\documentclass[11pt,a4paper]{article}
\pdfoutput=1
\usepackage{jinstpub}
\usepackage{subfigure}
\usepackage{url}
\usepackage[latin1]{inputenc} 
\usepackage{siunitx}
\usepackage[]{lineno}
\usepackage{xspace}
\usepackage{csquotes}

\newcommand{\TKpT}{\ensuremath{p_{_{\mathrm{T}}}}\xspace}

\newcommand{\cmsAuthorMark}[1]
{\hbox{\textsuperscript{\normalfont#1}}}

\title{
 Experimental Study of Different Silicon Sensor Options for the Upgrade of the CMS Outer Tracker
 }
\author[1]{The Tracker Group of the CMS Collaboration\note{Corresponding author: Georg Steinbr\"uck, 
Hamburg University, Germany}}
\emailAdd{georg.steinbrueck@desy.de}

\abstract{
During the high-luminosity phase of the LHC (HL-LHC), planned to start in 2027,   
the accelerator is expected to 
deliver an instantaneous peak luminosity of up to $7.5\times10^{34}$~cm$^{-2}$s$^{-1}$. 
A total integrated luminosity of $3000$ or even $4000$~fb$^{-1}$ is 
foreseen to be delivered to the general purpose detectors ATLAS and CMS over a decade, thereby increasing the discovery potential of the LHC experiments significantly.
The CMS detector will undergo a major upgrade for the HL-LHC, with entirely new tracking detectors consisting 
of an Outer Tracker and Inner Tracker. 
However, the new tracking system will be exposed to a significantly higher radiation  
than the current tracker, requiring new radiation-hard sensors.
CMS initiated an extensive irradiation and measurement campaign starting in 2009 
to systematically compare the properties of different silicon materials and design choices for the Outer Tracker sensors. 
Several test structures and sensors were designed and implemented on 18 different combinations of wafer 
materials, thicknesses, and production technologies.
The devices were electrically characterized before and after irradiation with neutrons, and with protons of different energies, with fluences  
corresponding to those expected at different radii of the CMS Outer Tracker after 
$3000$~fb$^{-1}$. 
The tests performed include studies with $\beta$ sources, lasers, and beam scans.
This paper compares the performance of different  options for the HL-LHC silicon sensors with a focus on silicon bulk material and thickness.
}


\begin{document}

\maketitle
\flushbottom

	




\section{Motivation}
Finely segmented silicon sensors are used in almost all high energy physics experiments for precision charged particle tracking.
Owing to their typical position close to the beam pipe they are subjected to high levels of irradiation by neutral 
and charged particles. Ionization in the oxide layer and at the interface to the silicon 
causes changes in the sensor properties. The main type of damage investigated in this paper, however,  is due to non-ionizing energy loss (NIEL) in the silicon  bulk. 
Defects with energy levels in the silicon band gap are created by removing silicon atoms from their lattice sites, 
thereby creating pairs of vacancies and interstitial atoms, which then lead to a number of energy levels due to defect kinetics. 
Depending on their properties, these energy levels  
have a threefold impact on basic silicon sensor characteristics~\cite{bib:HamburgModel, bib:RD50, bib:Moll}:
\begin{enumerate}  \itemsep 0 pt
\item Energy levels close to the middle of the band gap tend to increase the volume leakage current.
This raises the power consumption and heat load  of a sensor, and consequently increases the electrical noise.
\item Charged defects can modify the effective space charge concentration, which changes the voltage needed 
to achieve high-field regions in the entire volume of the sensor.  
Especially for high particle fluences, the electric field in the sensor bulk is altered from a linear dependence on depth to a double junction configuration~\cite{bib:Eremin}.
\item Some defects act as trapping centers for electrons or holes, thus reducing the amount of  charge collected.
\end{enumerate} 
%
The sensors in the current CMS microstrip tracker were produced from float-zone silicon wafers  using 
p$^+$ implants on n-type silicon (p-in-n). 
They were designed to withstand an integrated luminosity corresponding to 10 years of nominal LHC running 
($300$~fb$^{-1}$).
However, during the high-luminosity running phase of the LHC (from around 2026 onwards)~\cite{bib:sLHC, bib:Phase2TDR}, the instantaneous luminosity will be increased to  
5~$\times10^{34}$~cm$^{-2}$s$^{-1}$, or even to $7.5\times10^{34}$~cm$^{-2}$s$^{-1}$ in 
ultimate scenarios, with the goal of collecting an integrated luminosity of 3000 fb$^{-1}$ by the end of 2037. 
This corresponds to a 1 MeV neutron equivalent fluence~\cite{bib:HamburgModel, bib:Lindstroem1, bib:Lindstroem2} of $\phi_{\rm{eq}} = 9.4 \times 10^{14}$~cm$^{-2}$ at the innermost radius 
of the Outer Tracker (OT) ($r=\SI{22}{cm}$); 
therefore sensors optimized for these conditions are needed.

Silicon crystals produced with different growing techniques have been proposed and studied in the past~\cite{bib:RD50}. 
However, these sensors were produced for a variety of experiments by different vendors, and the corresponding measurements are often challenging to compare because the measurements 
were taken under different conditions. 
Therefore, starting in 2009, CMS embarked on an extensive campaign to systematically compare silicon materials, sensor designs, and layout parameters 
under otherwise identical conditions.
The same small silicon sensors and test structures for specific measurements were implemented by  one industrial supplier, 
Hamamatsu Photonics K.K. (HPK)~\cite{bib:hpk}, on 
a variety of silicon wafers differing in bulk material, thickness, production process, and whether the silicon bulk is n- 
or p-type (polarity).  
The main objectives of this paper are to address the issues of silicon bulk material and thickness, and to study individual strip parameters before and after irradiation.
The outcome of this extensive measurement program was used as input to the decision process for CMS OT sensor specifications. 

\section{The HL-LHC Upgrade of the CMS Outer Tracker}
\begin{figure}[t]
\begin{center}
\includegraphics[width=\textwidth]{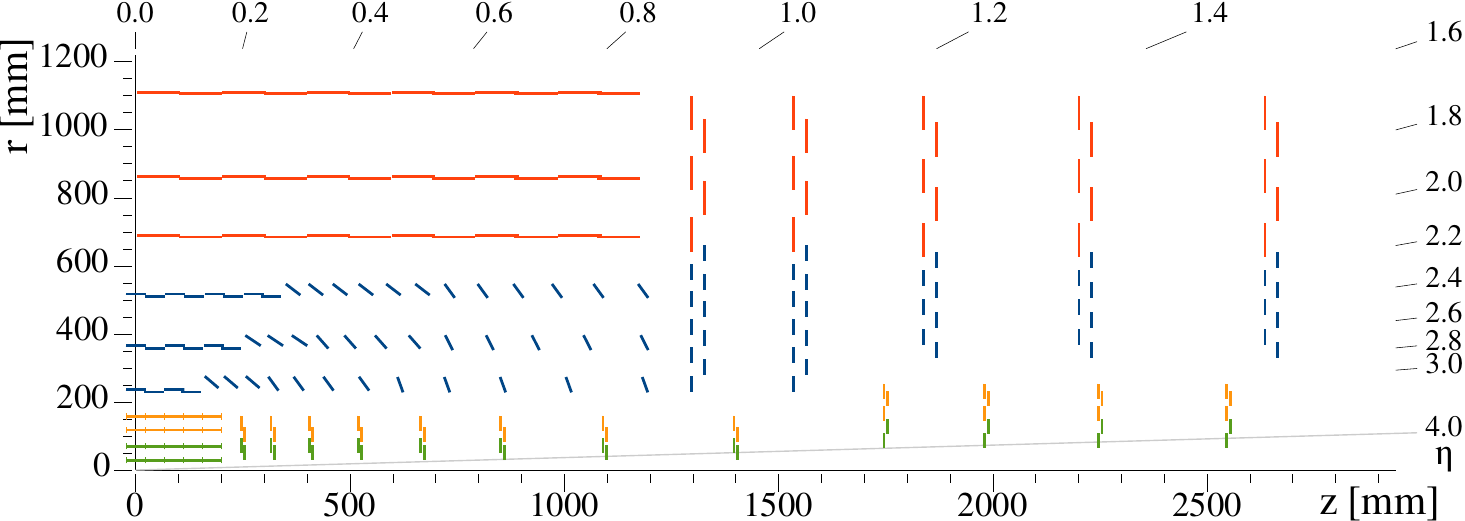}
\caption{
Sketch of one quarter of the CMS tracking system for the HL-LHC~\cite{bib:Phase2TDR} .
OT layers with modules consisting of two back-to-back strip sensors (2S modules)
are shown in red ($r~>~\SI{60}{cm}$), layers with modules consisting of a macro-pixel and a strip sensor (PS modules) are shown in blue ($\SI{20}{cm}~<~r~<\SI{60}{cm}$).  
The Inner Tracker, depicted in green and orange, consists of four pixel barrel layers ($r~<~\SI{20}{cm}$) and  12 disks per side.}
\label{fig:TrackerLayout}
\end{center}
\end{figure}

%
%
The design concept for the new CMS tracker~\cite{bib:Phase2TDR} is based on requirements to maintain efficient tracking capabilities under high-luminosity conditions.
With respect to the current tracker, the basic changes are an increased granularity to maintain hit occupancies in the percent range, 
a reduced material budget in the active region,
and delivery of tracker data to the Level 1 (L1) trigger
to significantly reduce the input rate to the high-level trigger.
Also, radiation tolerant silicon modules that will withstand 10 years of running at the HL-LHC are required. 
The overall baseline tracker design is shown in   Fig.~\ref{fig:TrackerLayout}.
To aid forward jet reconstruction in the presence of a high number of simultaneous proton-proton collisions at each bunch crossing (pileup), the 
angular coverage of the CMS tracking system is extended significantly 
up to a pseudorapidity of $|\eta| = 4.0$, mainly by adding  
forward pixel stations. This will significantly improve the physics performance in key processes such as  
vector boson fusion and vector boson scattering~\cite{bib:Phase2TDR}.

CMS~\cite{bib:cmsexp} adopts a right-handed coordinate system. The origin is centered at the nominal collision point inside the experiment. The $x$ axis points towards the center of the LHC, and the $y$ axis points vertically upwards. The $z$ axis points along the beam direction. The azimuthal angle, $\varphi$, is measured from the $x$ axis in the $x$-$y$ plane, and the radial coordinate in this plane is denoted by $r$. The polar angle, $\theta$, is measured from the $z$ axis. The pseudorapidity, $\eta$, is defined as $\eta =-\ln \tan{(\theta/2)}$. 
The momentum transverse to the beam direction, denoted by \TKpT, is computed from the $x$ and $y$ components.

The concept of the  OT is based on so-called \TKpT modules, which consist of two closely spaced silicon sensors 
for an on-module estimate of the 
transverse momentum of tracks, to be used in the L1 trigger. 
Hits on the two sensors, which are separated by \SI{1.6}{} to \SI{4}{mm} depending on the position of the module, are correlated in the front-end chip. This allows the discrimination of high- from low-\TKpT tracks based on their curvature in the \SI{3.8}{T} magnetic field of the CMS solenoid. Track ``stubs'' are formed from the selected correlated clusters in the inner  and the outer sensor. 
These track stubs are used off-detector to build tracks. 
A \TKpT module for the outer layers consists of two strip (2S) sensors,  while in the inner layers, owing to the higher track density, pixel-strip (PS) modules are used, consisting of a strip sensor and a macro-pixel sensor with  \SI{1.5}{mm} long macro-pixels.

The new tracker will use two-phase CO$_2$ cooling to remove the heat generated by the sensors and the electronics. 
This choice  makes it possible to operate and maintain the sensors at \SI{-20}{\celsius} or less,   
thus reducing the power consumption due to the leakage current of the sensors and  effectively preventing an increase in the  full depletion voltage, $V_{\rm fd}$, due to  reverse annealing. 
The material budget is also significantly reduced compared to mono-phase liquid cooling. Power losses on the low voltage cables will be reduced by using 
on-module DC-DC converters.
\section{Sensor Specifications for the CMS Outer Tracker}
The most important specifications for CMS OT sensors are listed  in Table~\ref{tab:sensorspecs}. They serve as guidelines for the measurements and results presented in this paper.
The figures before (after) irradiation refer to measurements at \SI{+20}{\degree C} (\SI{-20}{\degree C})  with relative humidity below 60\% (below 30\%).
\\
\begin{table}[!htb]
  \centering
  \caption{Selected requirements for CMS OT sensors.}
  \vskip 0.5cm
  \begin{tabular}{|c|c|c|}
    \hline
    
    \textbf{Parameter} & \textbf{Value for 2S}& \textbf{Value for PS}  \\
    \hline
    Depletion voltage*& \SI{< 350}{V} & \SI{< 180} - \SI{250}{V}\\
    Breakdown voltage& \SI{>800}{V} & \SI{>800}{V}\\
    Current density at \SI{600}{V}& \SI{\leq 2.5}{nA/mm^{3}} &\SI{\leq 5}{nA/mm^{3}} \\
    Sensor leakage current at \SI{600}{V}& \SI{\leq 7.25}{\micro A} &\SI{\leq 6}{\micro A}  \\
        \hline
    {  \bf Performance after irrradiation} &&  \\
     Target fluence & $\phi_{\rm eq} =3\times10^{14}$~cm$^{-2}$ & $\phi_{\rm eq} =1\times10^{15}$~cm$^{-2}$  \\
    Breakdown voltage at target fluence& \SI{>800}{V} & \SI{>800}{V}\\
     Sensor leakage current at \SI{600}{V} at target fluence& \SI{\leq 1}{m A} &\SI{\leq 1}{\m A}  \\
     Minimum signal at target fluence& \SI{>12000}{e} & \SI{>9600}{e}\\
    \hline
  \end{tabular}
  \label{tab:sensorspecs}
\end{table}
\linebreak
*Assuming sensor thicknesses of around \SI{300}{\micro\meter} 
    for 2S sensors, and  \SI{200}-\SI{240}{\micro\meter} for PS sensors.

\section{Structures and Wafer Layout}
\begin{figure}[t]
\begin{center}
\includegraphics[width=0.5\textwidth, angle=0]{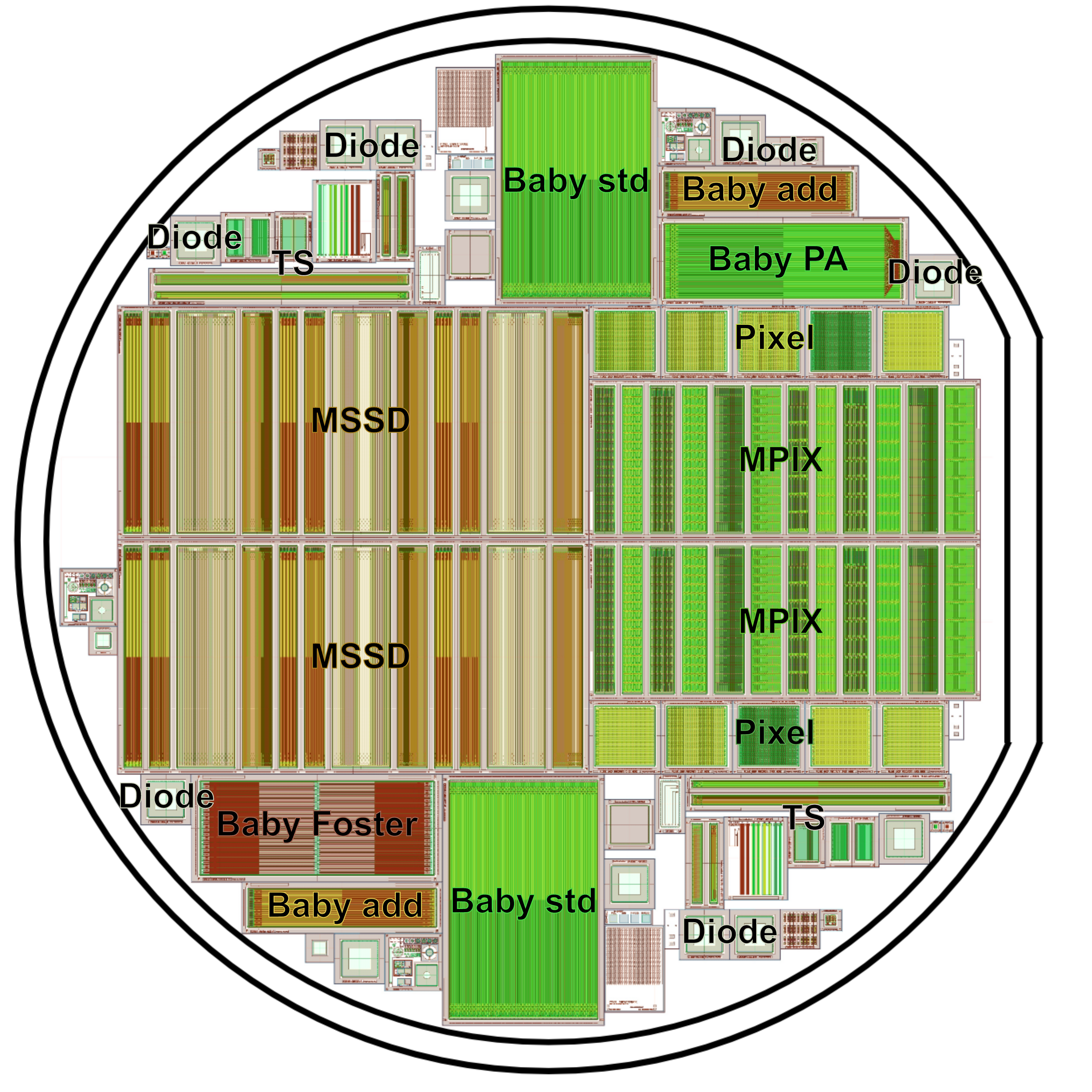}
\caption{Layout of the wafer with 28 different structures, 
including pad diodes, small strip sensors (Baby std, Baby add),
a baby strip sensor with pitch adaptor (Baby PA), test structures (TS)
containing  a MOS structure, a diode, a capacitance structure, a sheet resistance structure,  
a special FOurfold segmented STrip sensor with Edge Readout (FOSTER),
a Multi-geometry Silicon Strip Detector (MSSD), 
and a Multi-geometry Pixel sensor (MPix).}
\label{fig:Wafer}
\end{center}
\end{figure}

\begin{figure}[t]
\begin{center}
\includegraphics[width=0.7\textwidth, angle=0]{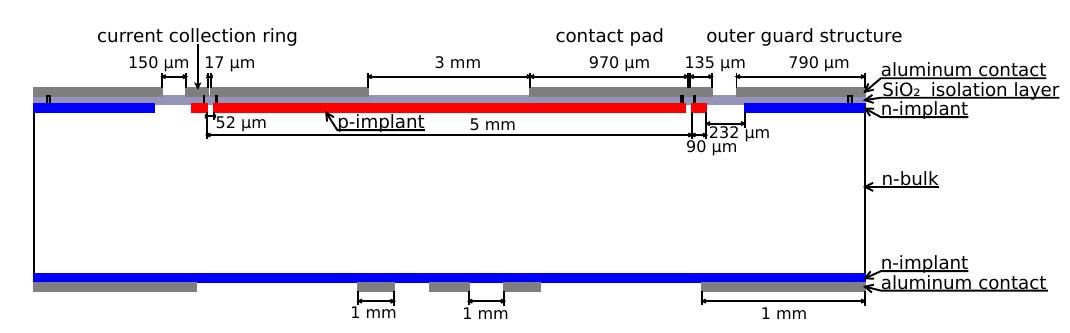}
\caption{Cross section of a p-in-n diode (not to scale).
The aluminum contact on the junction side has a \SI{3}{mm} wide opening; 
on the backside, a \SI{1}{mm} wide aluminum grid allows for penetration of laser light.
}
\label{fig:diode}
\end{center}
\end{figure}

\begin{figure}[t]
\begin{center}
\includegraphics[width=0.7\textwidth]{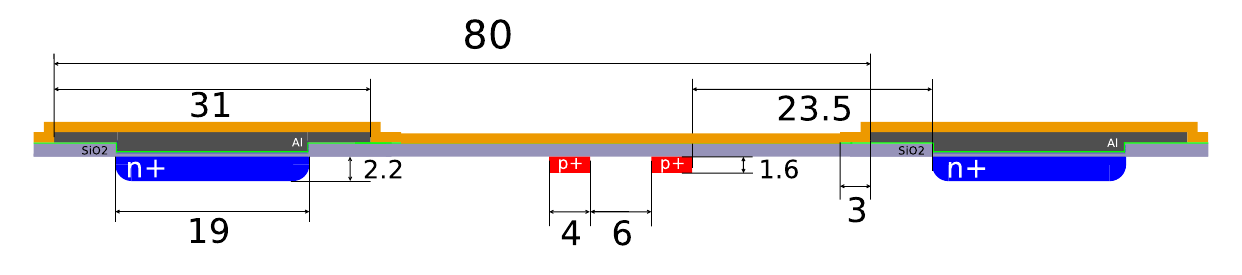}
\caption{Cross section of the segmented side of an n-in-p mini-strip sensor with p-stop strip isolation. 
All lengths are in units of \SI{}{\micro\meter}. The passivation layer is drawn in orange. 
With exception of the p-stop implants, this layout also applies to the n-in-p sensors with p-spray isolation, and to the p-in-n sensors.}
\label{fig:stripCS}
\end{center}
\end{figure}

To study the properties of the different materials and production processes before and after irradiation, 
28 different structures were specifically designed for this project run (Fig.~\ref{fig:Wafer}).
Several fully functional strip and pixel sensors were implemented. 
Large areas of the wafer are devoted to structures with different strip and pixel geometries. 
The remaining part of the wafer was covered with specialized test structures, which give access to parameters 
that cannot be measured with a sensor.
The structures investigated for this paper are \SI{5}{\milli\meter }$~\times$~\SI{5}{\milli\meter} pad diodes to study basic material properties 
(Fig.~\ref{fig:diode})  
and AC coupled mini-strip sensors with \SI{80}{\micro\meter} pitch and a length of either \SI{3.27}{\cm} or \SI{2.57}{\cm} 
(Fig.~\ref{fig:stripCS}). 
The design parameters of these mini-strip sensors, summarized in Table~\ref{tab:sensor}, 
are similar to those for the sensors used in the current CMS OT. 
%
%

\begin{table}[!htbp]
  \centering
  \caption{Layout and process details of the mini-strip sensors~\cite{bib:ptype}. There are two versions: one with 256 strips and \SI{3.27}{cm} strip length, and one with 64 strips and \SI{2.57}{cm} strip length.
With the exception of the parameters concerning the p-stop implants and the p-spray, the parameters apply both for the n-in-p and p-in-n sensors. 
  }
\vskip 0.5 cm
  \begin{tabular}{|c|c|}
    \hline
    \textbf{Parameter} & \textbf{Value} \\
    \hline
    Strip length & \SI{3.27}{\centi\meter} and \SI{2.57}{\centi\meter}\\
    Strip width & \SI{19}{\micro\meter}\\
    Strip pitch & \SI{80}{\micro\meter}\\
    Metal overhang & \SI{6}{\micro\meter}\\
    Number of strips & 256 and 64\\
    Overall dimensions & \SI{3.5}{\centi\meter} $\times$ \SI{2.3}{\centi\meter}  and  \SI{2.8}{\centi\meter} $\times$ \SI{1.3}{\centi\meter}\\
    Coupling dielectric thickness & \SI{300}{\nano\meter}\\
    Strip doping concentration (peak) & $\sim$ \SI{1e19}{\centi\meter^{-3}}\\
    Strip implant depth & \SI{2.2}{\micro\meter}\\
    p-stop doping concentration (peak/ integrated) & $\sim$ \SI{5e15}{\centi\meter^{-3}}/ $\sim$ \SI{2e11}{\centi\meter^{-2}}\\
    p-stop  depth & $\sim$\SI{1}{\micro\meter}\\
    p-spray doping concentration (peak/ integrated) & $\sim$ \SI{1e15}{\centi\meter^{-3}}/ $\sim$ \SI{5e10}{\centi\meter^{-2}}\\
    p-spray  depth &  $\sim$\SI{1}{\micro\meter}\\
    \hline
  \end{tabular}
  \label{tab:sensor}
\end{table}

\section{Materials, Thicknesses, and Production Technologies}
\begin{figure}[t]
\begin{center}
\includegraphics[width=0.8\textwidth]{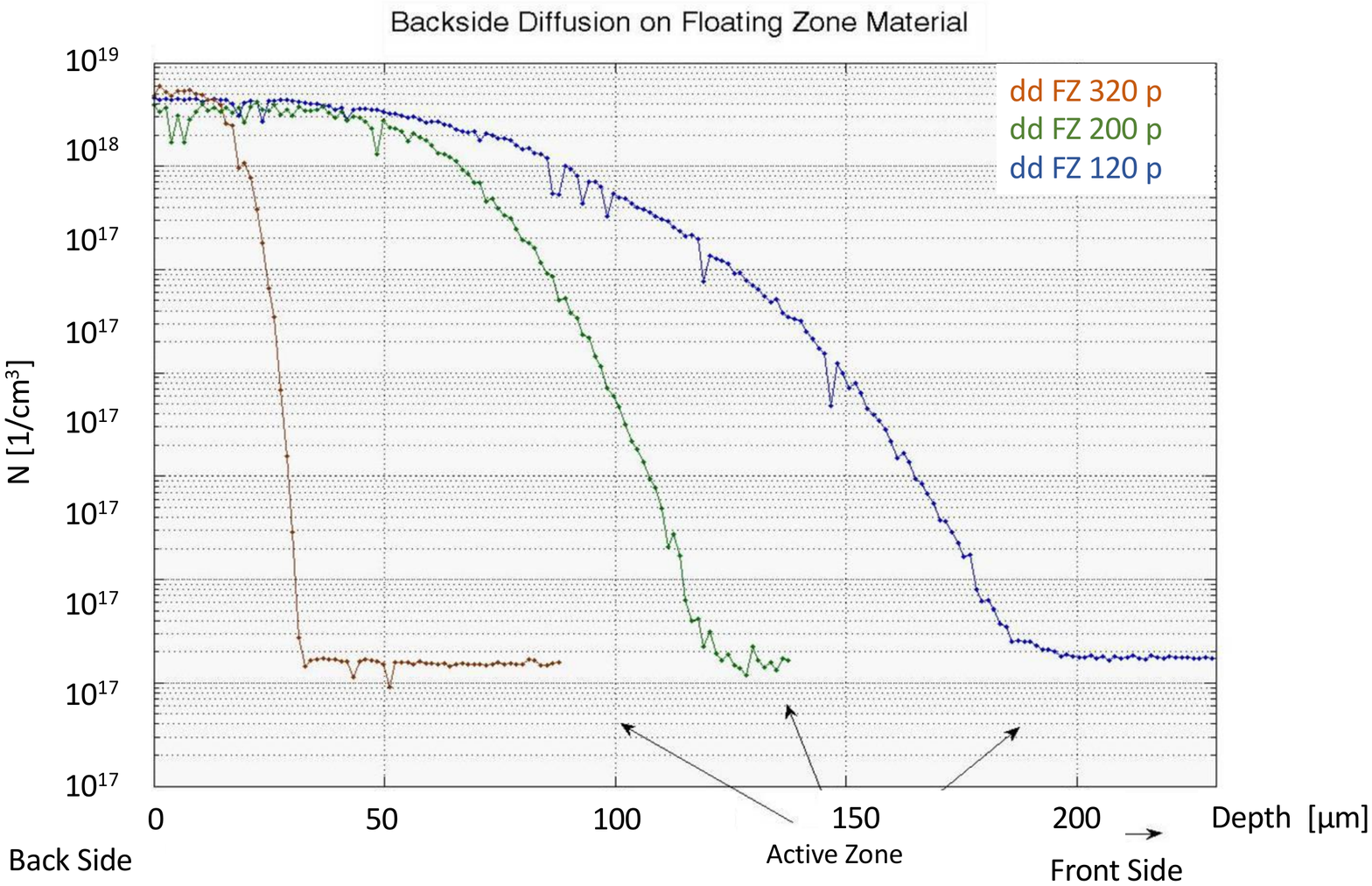}
\caption{Doping profile for deep diffused p-type sensors obtained by Spreading Resistance Profiling (SRP).  
}
\label{fig:dd}
\end{center}
\end{figure}

Two important silicon wafer parameters that determine the evolution of full depletion voltage, signal, and noise with fluence are the active thickness and the oxygen content. 
The selected materials and thicknesses cover the relevant combination of parameters (Table~\ref{tab:materials}), including the standard \SI{320}{\micro\meter} float-zone (FZ) material as a reference. 
In the current CMS strip tracker \SI{320}{} and \SI{500}{\micro\meter} thick float-zone sensors are used.
All materials studied here have <100> crystal orientation.
\begin{table}[!htbp]
\begin{center}
\caption{Wafer materials and thicknesses studied for the CMS tracker upgrade. 
The sensors were produced both on n-type and p-type bulk silicon for each combination of material and thickness listed. 
The physical thickness has a variation of less than \SI{\pm10}{\micro\meter} according to specifications.
Sensors fabricated on dd-FZ 300 are also labeled as
dd-FZ-320 and FZ320 in some plots.
FZ: float-zone silicon, MCz: magnetic Czochralski silicon, dd: deep diffusion on the wafer backside to achieve the desired active thickness on \SI{320}{\micro\meter} wafers.}
\vskip 0.5 cm
\begin{tabular}{|p{1.55cm}|p{0.6cm}|p{0.6cm}|p{0.6cm}|}
\hline
 & \multicolumn{3}{|c|}{Active thickness (\SI{}{\micro\meter})}\\
 \cline{2-4}
Material &  300 & 200 & 120\\
\hline
FZ  & - & X & X \\
\hline
dd-FZ & X & X & X \\
\hline
MCz   & - & X & - \\
\hline
\end{tabular}
\label{tab:materials}
\end{center}
\end{table}%
For the materials denoted with \emph{dd} the active thickness is reduced by a special treatment provided by HPK  (\emph{deep diffusion}). 
The physical thickness of these wafers is always  
$320~\SI{~\pm~10}{\micro\meter}$ while the active thickness is reduced to the desired value by the diffusion of dopants from the backside.  
A doping profile for deep diffused p-type sensors obtained by Spreading Resistance Profiling (SRP) is shown in Fig.~\ref{fig:dd}.
Compared to direct-wafer bonded silicon, the deep diffusion process leads to a more gradual transition from the low resistivity back side to the high resistivity bulk.  
This highly doped backside serves effectively as an ohmic contact to the remaining active region. The active thickness relevant for full depletion, electric field, and charge collection is thereby 
decoupled from the sensor thickness, which impacts mechanical stability and thermal performance. 
Moreover, for mechanical stability, very thin wafers need to be bonded to a carrier wafer during processing, an additional step that 
is not needed for the wafers treated with the deep diffusion process. 
The differences in the electrical properties of sensors produced using wafers processed with these two techniques are studied before and after irradiation. 
In addition, \SI{200}{\micro\meter} thick magnetic Czochralski (MCz) silicon sensors are studied. 
MCz has the advantage of a high oxygen concentration introduced during the crystal growth process in a quartz (SiO$_2$) crucible.

The structures on all the selected materials were manufactured in three different processes: standard p-in-n, n-in-p with p-stop strip isolation, 
and n-in-p with p-spray strip isolation.
In the current CMS strip tracker, p-in-n sensors are used. While initially both p-in-n and n-in-p options were considered for the upgraded CMS OT at the HL-LHC,
the observation of non-Gaussian noise in p-in-n prototype sensors led to the decision to use n-in-p sensors~\cite{bib:ptype}.
Moreover, n-in-p sensors have the additional benefit
that they always deplete from the readout side, which leads to good signal collection even for heavily irradiated sensors that are not fully depleted anymore.
A drawback of n-in-p sensors is the need for p-stop or p-spray strip isolation to interrupt the electron inversion layer, which forms as a result of positive oxide charges.

The boron and phosphorus bulk doping concentrations are around $3\times10^{12}$~cm$^{-3}$ for the float-zone and 4--$5\times10^{12}$~cm$^{-3}$ for the  MCz sensors .
The doping concentration parameters measured by  spreading resistance profiling (SRP)~\cite{bib:srp}  
are listed in Table~\ref{tab:sensor}. SRP is a technique used to measure resistivity as a function of depth in semiconductors.

\begin{figure}[t]
\begin{center}
\includegraphics[width=0.5\textwidth]{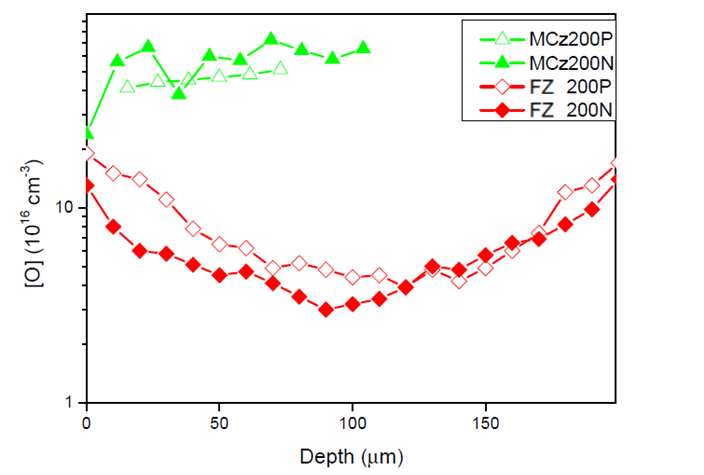}
\caption{Oxygen concentration [O] as a function of depth for 
\SI{200}{\micro\meter} thick float-zone (FZ) and 
magnetic Czochralski (MCz) n-in-p (P) and p-in-n (N) pad diodes. The depth is measured from the junction side of the diode. Varying much less with depth, 
the measurements for the MCz diodes were only performed up to a depth of either 75 or \SI{100}{\micro\meter}.}
\label{fig:Oxygen}
\end{center}
\end{figure}

\begin{figure}[t]
\begin{center}
\begin{minipage}{.49\textwidth}
\includegraphics[width=1.1\linewidth]{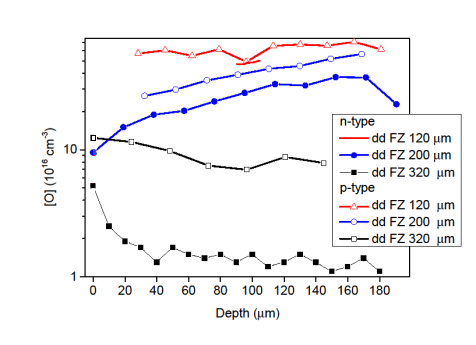}
\end{minipage}
\hfill
\begin{minipage}{.49\textwidth}
\includegraphics[width=0.87\linewidth]{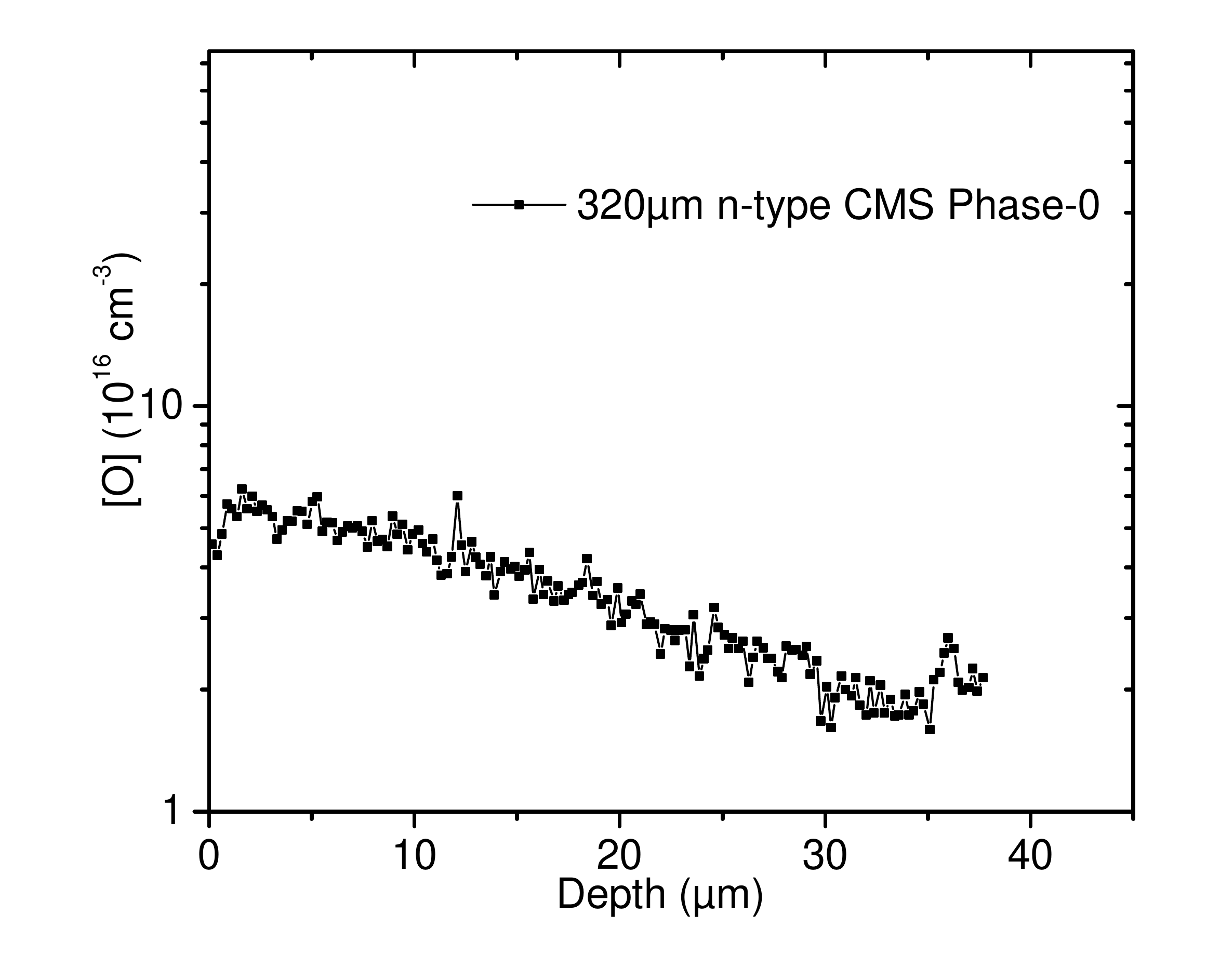}
\vspace {-0.15 cm}
\end{minipage}
\caption{Oxygen concentration [O] as a function of depth for 
deep diffused (dd) float-zone pad diodes (left).  The depth is measured from the junction side of the diode. For the n-type sensor with \SI{120}{\micro\meter}  active thickness only one 
data point at a depth of \SI{95}{\micro\meter}  was available.
For comparison, the oxygen concentration for a \SI{320}{\micro\meter} thick HPK n-type float-zone sensor as used in the current CMS tracker (Phase-0) is shown (right).
It was measured up to a depth of \SI{38}{\micro\meter}.
}
\label{fig:OxygenFZ}
\end{center}
\end{figure}

The oxygen concentration was measured using the secondary ion mass 
spectrometry (SIMS) technique, the results of which are shown in Figs.~\ref{fig:Oxygen} 
and ~\ref{fig:OxygenFZ}. The oxygen concentration ranges from $1\times10^{16}$~cm$^{-3}$ for the \SI{320}{\micro\meter}  
thick p-in-n float-zone material to $5\times10^{17}$~cm$^{-3}$ for the MCz material, and varies with depth. 
All materials studied (except for the n-type dd-FZ 320) are rather oxygen-rich compared to standard float-zone 
silicon, with  typical oxygen concentrations in the range 0.5--$1\times10^{16}$~cm$^{-3}$. 
The reason for the relatively large difference in oxygen concentration between the  \SI{320}{\micro\meter}  thick deep diffused n and p type material is unknown.
The oxygen concentration of the sensors in the current CMS OT is  ${\approx}2\times10^{16}$~cm$^{-3}$ (Fig.~\ref{fig:OxygenFZ}, right plot). 
These sensors have a deep diffused ohmic backside contact like in the dd-FZ 320 sensors studied in this paper.
%
%
%
%

\section{Irradiation Campaign}
~\label{sec:irrad}
\begin{figure}[t]
\begin{center}
\includegraphics[width=0.7\textwidth]{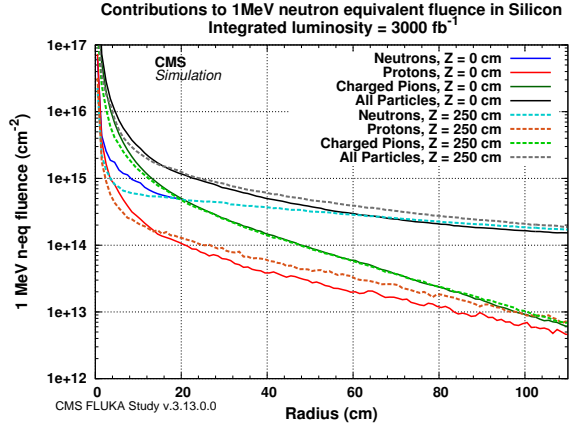}
\vskip -1.2 cm
\caption{Fluence versus radius expected at CMS for an integrated luminosity of $3000$~fb$^{-1}$ at the HL-LHC, based on FLUKA simulations~\cite{bib:FLUKA1, bib:FLUKA2}.
The fluence is shown for the central part of the upgraded tracker ($z$ = \SI{0}{cm}), and for the very forward region  ($z$ = \SI{250}{cm})~\cite{bib:fluenceplot}.
}
\label{fig:Fluenceplot}
\end{center}
\end{figure}

All structures were first electrically characterized, after which they were irradiated with reactor neutrons or proton beams to 
different fluences in several steps. After each irradiation step, the structures were annealed and again electrically characterized. The structures 
initially exposed to neutron irradiation later received a proton irradiation and vice versa, followed by another annealing 
treatment and electric characterization. This strategy allows us to investigate the properties of the materials after pure 
proton and neutron irradiation, and with mixed irradiation, using a minimum number of material samples.  
The fluence for charged and neutral hadrons at CMS, estimated with FLUKA simulations~\cite{bib:FLUKA1, bib:FLUKA2} for an integrated luminosity of $3000$~fb$^{-1}$ at the HL-LHC, is 
shown in Fig.~\ref{fig:Fluenceplot} as a function of the radial distance from  the beam.
Table~\ref{tab:irradiations} shows the fluence steps for neutron and proton irradiations corresponding to different radii 
of the OT of CMS as chosen for this study.
The samples were irradiated with neutrons at the TRIGA Mark II reactor~\cite{bib:Triga} at the 
Josef-Stefan-Institute in Ljubljana, Slovenia, corresponding to a 1 MeV equivalent hardness factor of 0.9~\cite{bib:GregorThesis}.
To study the dependence of 
radiation damage effects on the proton energy, several facilities were utilized for proton irradiation:
the Karlsruhe Compact Cyclotron (KAZ) (23 MeV)~\cite{bib:kaz} operated by the ZAG Zyklotron AG, the Proton Radiography Facility (pRad) at the Los Alamos LANSCE accelerator facility (800~MeV)~\cite{bib:losAla}, and the CERN Proton Synchrotron 
(\SI{23}{GeV})~\cite{bib:ps}, where the values represent the kinetic energy of the protons. 
In this paper results for 
samples irradiated with reactor neutrons and \SI{23}{GeV}    and \SI{23}{MeV}  protons are shown.
The measured hardness factors are  \num[separate-uncertainty = true]{0.62\pm 0.04} for \SI{23}{GeV}  and \num[separate-uncertainty = true]{2.20\pm 0.43}  for \SI{23}{MeV} protons~\cite{bib:hardness}.
%
%
\begin{table}[t]
\begin{center}
\caption{Fluences (in \SI{1}{\MeV} neutron equivalent) chosen in the irradiation study for different radii of the 
tracker of CMS. The last column 
indicates that thicker sensors are studied for the outer radii while thin silicon is considered for the innermost region (especially the pixel layers).
The fluences are calculated for an integrated luminosity of 3000~fb$^{-1}$ 
and are inflated by 50$\%$ to accommodate uncertainties in the FLUKA~\cite{bib:FLUKA1, bib:FLUKA2} simulations 
and the potential delivery of additional luminosity in ultimate HL-LHC 
luminosity scenarios.}
\vskip 0.5 cm
\begin{tabular}{|p{1.2cm}|p{1.8cm}|p{1.8cm}|p{1.8cm}|p{1.8cm}|}
\hline
Radius [cm]& Proton $\phi_{\rm eq}$[cm$^{-2}$] & Neutron $\phi_{\rm eq}$[cm$^{-2}$] & Total $\phi_{\rm eq}$[cm$^{-2}$]  & Active thickness \\
\hline
\hline
40  & $3\times 10^{14}$ & $4\times 10^{14}$ & $7\times 10^{14}$ & $\geq$~\SI{200}{\micro\meter} \\
\hline
20  & $1\times 10^{15}$ & $5\times 10^{14}$ & $1.5\times 10^{15}$ & $\geq$~\SI{200}{\micro\meter} \\
\hline
15  & $1.5\times 10^{15}$ & $6\times 10^{14}$ & $2.1\times 10^{15}$ &  $\leq$~\SI{200}{\micro\meter} \\
\hline
10  & $3\times 10^{15}$ & $7\times 10^{14}$ & $3.7\times 10^{15}$ & $\leq$~\SI{200}{\micro\meter} \\
\hline
5  & $1.3\times 10^{16}$ & $1\times 10^{15}$ & $1.4\times 10^{16}$ & $<$~\SI{200}{\micro\meter} \\
\hline
\end{tabular}
\label{tab:irradiations}
\end{center}
\end{table}%
%
After irradiation, the sensors were subjected to annealing in several steps as shown in Tab.~\ref{tab:annealing}.
The annealing was done at  \SI{60}{\degree C}, and at \SI{80}{\degree C} for the latter steps to avoid very long annealing times.
The data shown in this paper are typically presented scaled to equivalent annealing times at a reference temperature of \SI{21}{\degree C} (room temperature) or \SI{60}{\degree C}.
The scaling is based on a parametrization of the current related damage rate $\alpha$ as a function of annealing time at different temperatures from Ref.~\cite{bib:Moll}.
For a given annealing time and temperature, an $\alpha$  value is determined, and then the equivalent time at the reference temperature is obtained as the time leading to the same 
$\alpha$.
%
\begin{table}[t]
\begin{center}
\caption{Temperatures and time steps used in annealing study. The indicated times are the individual steps, which have to be added to obtain the total annealing time.}
\vskip 0.5 cm
\begin{tabular}{|p{2.0cm}|p{1.2cm}|p{1.2cm}|p{1.2cm}|p{1.2cm}|p{1.2cm}|p{1.2cm}|p{1.2cm}|}
\hline
Temperature & \SI{60}{\degree C}	&\SI{60}{\degree C}	&\SI{60}{\degree C}	&\SI{60}{\degree C}&\SI{80}{\degree C}	&\SI{80}{\degree C}	&\SI{80}{\degree C}\\
\hline
Annealing time & 20 min	&20 min	&40 min	&76 min	&15 min	&30 min	&60 min\\
\hline
\end{tabular}
\label{tab:annealing}
\end{center}
\end{table}%


\section{Measurement Techniques}
\subsection{Determination of Full Depletion Voltage and Leakage Current}
%
%
The test sensors (pad diodes and mini-strip sensors) are characterized in a current-voltage ($I$-$V$) 
and capacitance-voltage ($C$-$V$) measurement setup that allows the sensors to be cooled down to \SI{-30}{\degree C}. 
A single guard ring surrounding the pad sensor is used to 
contain the electric field in the active volume to 
ensure high voltage stability and reduce the leakage current (Fig.~\ref{fig:diode}).
Unless stated otherwise, the guard ring of the pad sensor under test is grounded to ensure a well-defined sensitive volume.
The strip sensors are surrounded by a bias ring, which is grounded during operation, and an outer guard ring, which is left floating. 
The humidity is reduced by a constant  flow of dry air.

The full depletion voltage is extracted from $C$-$V$ measurements, where the capacitance is evaluated in parallel mode.
For non-irradiated sensors, the capacitance decreases with voltage  
and reaches a minimum (geometrical capacitance) when the sensor is fully depleted.
For heavily irradiated sensors, the interpretation of the measurement results is more complicated.
The measured capacitance strongly depends on the measurement temperature and frequency  
since irradiation-induced trap sites are filled and depleted by the AC voltage applied in the measurement. 
Whether a certain defect contributes depends on the characteristic time scale for filling and emitting in comparison with the inverse of the measurement frequency. The emission and capture probabilities for electrons and holes for each defect level are temperature dependent. 
%
%
%
Capacitance-voltage measurements were performed at \SI{0}{\degree C} at 1 kHz, 
and at \SI{-20}{\degree C} at 1 kHz and 455 Hz.
The full depletion voltage is determined from \SI{0}{\degree C} measurements at 1kHz  and  for \SI{-20}{\degree C} measurements at 455 Hz.
For the full depletion voltage the value is used at which the linear rise of $1/C^2$ versus voltage reaches an approximate  plateau,
by fitting straight lines to the data points below and above the kink, excluding the transition region, and determining their intersection.
A low frequency like 455 Hz ensures that traps can dynamically be charged and discharged, given that emission timescales are in the millisecond-range at   \SI{-20}{\degree C}.
If not stated otherwise, the full depletion voltage results for irradiated sensors shown in this paper are obtained from $C$-$V$ measurements 
at  \SI{-20}{\degree C}  and 455 Hz.  
As an alternative approach, recently, models with a position-dependent resistivity have been used to describe the frequency dependence of the parallel capacitance for irradiated sensors~\cite{bib:PositionDependent}.
%
%
%
%
%
\subsection{Measurements of Charge Collection for Diodes}
\begin{figure}[t]
\begin{center}
\includegraphics[width=0.5\textwidth]{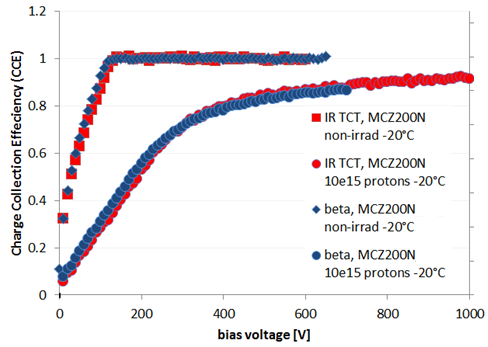}
\caption{Charge collection efficiency ($CCE$) measured with an infrared laser (red symbols) and a $\beta$ source (blue symbols).
The  curves shown refer to a non-irradiated diode and to a diode irradiated with protons to $\phi_{\rm eq} = 10^{15}$~cm$^{-2}$.
}
\label{fig:CCEcomp}
\end{center}
\end{figure}

The reduction of the collected charge due to trapping of charge carriers 
by radiation-induced defect levels in the silicon band gap is one of the main concerns for the HL-LHC tracker.
The charge collection efficiency ($CCE$) at a certain bias voltage, $V$, is 
defined as the fraction of the signal measured for an irradiated diode at $V$  
compared to that of a non-irradiated diode of the same type measured 
at a reference voltage of \SI{400}{V}. 
Charge carriers for this measurement can be generated either using an infrared laser or 
a $\beta$ source. While the latter provides a well-defined mean energy deposited in the silicon, the advantage of the 
laser is that the signal can be increased well above the noise level 
and the position of the laser light can be well controlled.
Figure~\ref{fig:CCEcomp} shows the $CCE$ as a function of bias voltage for a non-irradiated and an irradiated diode, 
measured each with a laser and with a $\beta$ source (Strontium-90). 
It can be seen that the measurements agree rather well.
%
%
%
%
%
For the $CCE$ measurement with the laser, a {transient current technique} (TCT) setup~\cite{bib:TCT} is used
with an infrared laser with 1063 nm wavelength and a pulse width of \SI{50}{ps} (FWHM).
The absorption length  at this wavelength is of the order of \SI{1}{\milli\meter}, 
significantly larger than the typical thickness of  a silicon sensor, 
which leads to an approximately uniform electron-hole pair creation as a function of depth, similar 
to the profile of a minimum ionizing particle.
%
%
The resulting signal is amplified with a fast current preamplifier and recorded with a GHz
bandwith digital oscilloscope.
%
The charge collection efficiency is obtained by integrating the pulse over time 
and comparing the total charge (in arbitrary units) to the one obtained for a 
non-irradiated reference sensor at a reference voltage (\SI{400}{V}) that is above the full depletion voltage.  
A stability of the measured charge collection efficiency over time of better than $3 \%$ can be reached in this kind of measurement. 
The uncertainty is dominated by the stability of the laser pulse. 
%
The intensity of the laser is set high enough to obtain a very good signal-to-noise ratio, but well below the onset of the "plasma effect"~\cite{bib:plasma}. This was ensured by requiring the charge deposited by a single 
laser pulse to be less than 50 times that induced by a minimum ionizing particle.
Moreover, the laser light is not focussed and has a spot size of about \SI{300}{\micro\meter} at the sensor surface.
Fluctuations are further reduced by averaging signals over 512 events. 
%
%
%
\subsection{Measurement of Charge Collection for Strip Detectors with the ALIBAVA Setup}
\begin{figure}[t]
\begin{center}
\includegraphics[width=0.5\textwidth]{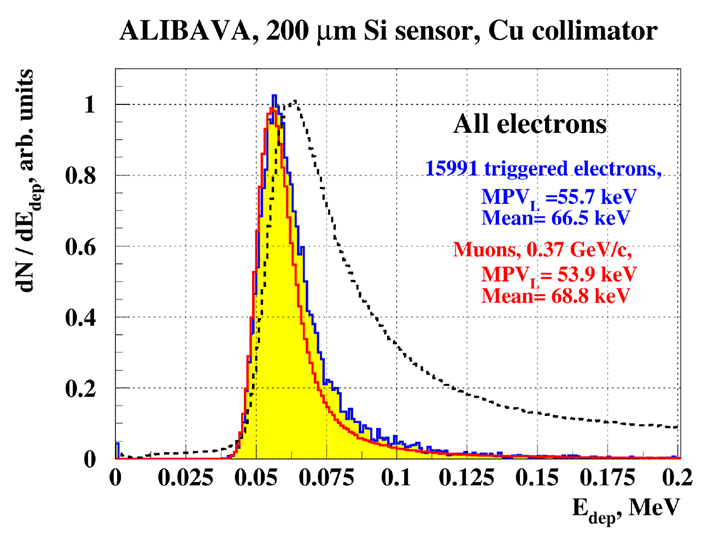}
\caption{GEANT3 simulation  of the $\beta$ test stand~\cite{bib:erflethesis}. The deposited energy, $E_{\rm dep}$, is shown for all electrons (black), and  
for electrons triggered by the scintillators (blue).
For comparison, the energy deposited by muons with a momentum of 0.37 GeV is plotted (red).
The distributions are fitted by a Landau distribution convolved with a Gaussian (fit not shown for readability). 
The most probable value of the Landau (denoted as MPV$_{\rm L}$) 
and the mean of the distributions are given.}
\label{fig:electrons}
\end{center}
\end{figure}

For the measurement of charge collection in strip sensors {\it A Liverpool Barcelona Valencia} (ALIBAVA) setup~\cite{bib:alibava, bib:alibava2} is used, based on the LHCb Beetle readout chip.
Charge is generated using an infrared laser (wavelength \SI{1063}{\nano\meter}) or a $\beta$ source (Strontium-90). 
In the first case, a signal from the laser driver is used to trigger the readout of the data. 
In case of the $\beta$ source, a trigger based on one or two plastic scintillator planes on the side opposite to the source is used, 
thus imposing an energy cut on the electrons.
With this energy cut,
both shape and normalization of the distribution of electron-hole pairs generated by electrons 
in \SI{200}{\micro\meter} of  silicon have been shown to be very similar to that of a minimum ionizing particle 
(Fig.~\ref{fig:electrons})~\cite{bib:erflethesis}.
The settings of the pre-amplifier and shaper of the Beetle chip can be adjusted to obtain fast pulse shapes compatible with the 40 MHz clock of the LHC.
However, when used with a $\beta$ source, which is non-synchronous with the Beetle clock,  a longer pulse shape with a peak region of about 20 ns was used to ensure that a large fraction of the triggered events is useable for analysis. The peak region is defined as the part of the pulse which is within $90 \%$ of the maximum pulse height. 
While the shaping time has an impact on the strip noise for irradiated sensors with substantial leakage current, charge collection measurements are less impacted.
%

\subsection{Measurements of Strip Parameters with a Probe Station}
This  description was originally published in Ref.~\cite{bib:ptype} and is repeated here for completeness. 
Initially, all sensors were electrically characterized with a probe station measuring the following quantities.
\begin{itemize}
\item {\bf Total leakage current:}
The current in the bias line is measured versus bias voltage ($I$-$V$) with
floating guard ring. 
The HPK sensors typically had current densities lower than \SI{2}{nA/mm^3}.
\item {\bf Total capacitance:}
The capacitance of the sensor is measured versus bias voltage ($C$-$V$) with floating
guard ring to extract the full depletion voltage.
\item {\bf Strip leakage currents:}
The leakage currents of individual strips are measured to check the uniformity.

\item{\bf Coupling capacitance:} The capacitance between strip implant and metal strip is measured at
100 Hz and should be larger than 1.2 pF$/$cm per $\upmu$m of implanted strip width.

\item{\bf Current through the dielectric:} The current is measured between strip implant and metal strip
applying 10 V and should be smaller than 1 nA.

\item {\bf Bias resistance:} The bias resistor at each strip is evaluated by measuring the current when applying
2V to the DC pad. A resistance between 1 and \SI{3}{M\Omega} is envisaged.

\item {\bf Interstrip capacitance:} The capacitance between neighboring metal strips per unit strip length is measured at 1 MHz
and should be below 1 pF/cm.


\item {\bf Interstrip resistance:} The resistance between two strip implants is evaluated by measuring the $I$-$V$
characteristic from \SI{-1}{V} to \SI{1}{V}. It should be ten times higher than the bias
resistance, and the resistivity should be larger than \SI{10}{G\Omega\cdot cm} before irradiation.

\end{itemize}   
%

\section{Results}
\subsection{Leakage Current}
One of the factors compromising the performance of a tracking detector over time is the increased sensor leakage current due to radiation damage.
The consequences are increased electrical noise and power consumption, which increases the heat load and determines the temperature at which the sensors need to be operated to avoid thermal runaway. 
Thermal runaway describes a  situation in which the cooling power is insufficient to cool the module, and the sensor temperature rises in
a positive feedback loop, since the leakage current grows exponentially with temperature.
The bias voltage, leakage current, and power consumed by the sensors impact 
the choice of power supplies and the design of the cooling system.
%
%
On the other hand, technical limitations, such as maximum supply voltages or the lowest achievable sensor temperatures, may have an influence on the operating parameters and even the design of the sensors.

\begin{figure}[t]
\begin{center}
\includegraphics[width=0.5\textwidth]{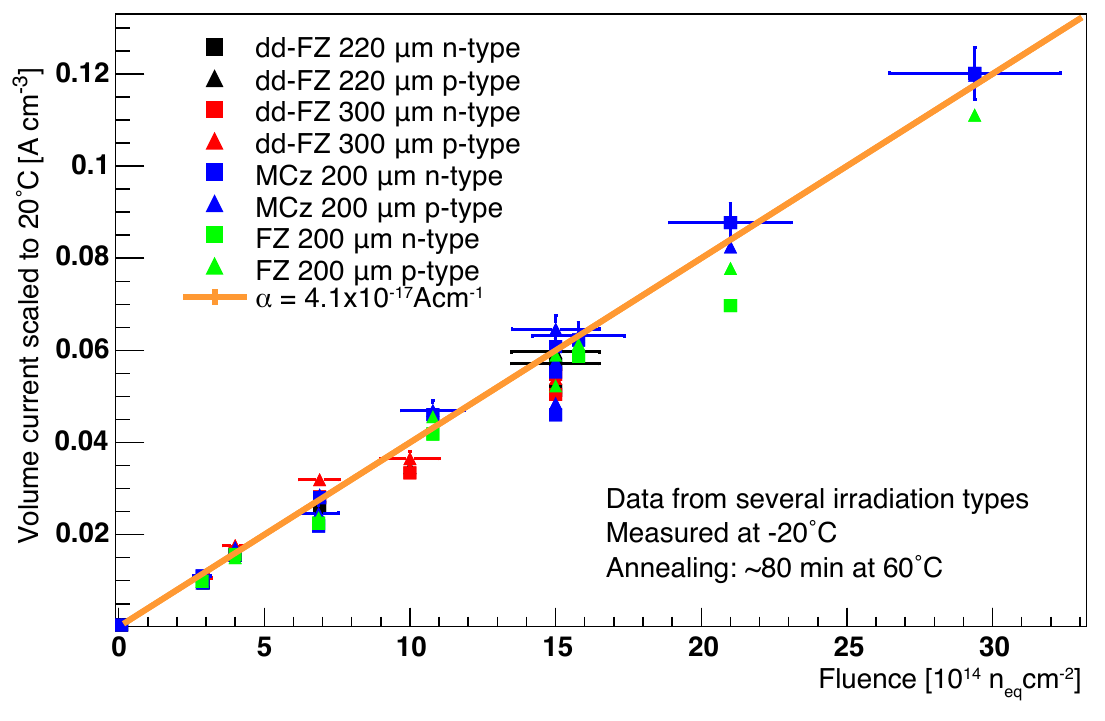}
\caption{Volume current density as a function of  \SI{1}{\MeV} neutron equivalent fluence measured at \SI{-20}{\degree C}  and scaled to  \SI{+20}{\degree C}.
For comparison, the proportional dependence of the volume current on fluence based 
on a previous fit to a variety of measurements~\cite{bib:Moll} is depicted by a solid line.
The uncertainty in the fluence is estimated to be about 10$\%$, the estimated uncertainty in the currents is 
mainly due to uncertainties in the determination of the full depletion voltage, the measurement temperature, and the active thickness
of the diodes, especially for dd-FZ. For improved  readability, 
the uncertainties are omitted for some points as they are all of similar size.}
\label{fig:Current}
\end{center}
\end{figure}

\begin{figure}[t]
\begin{center}
\includegraphics[width=0.5\textwidth]{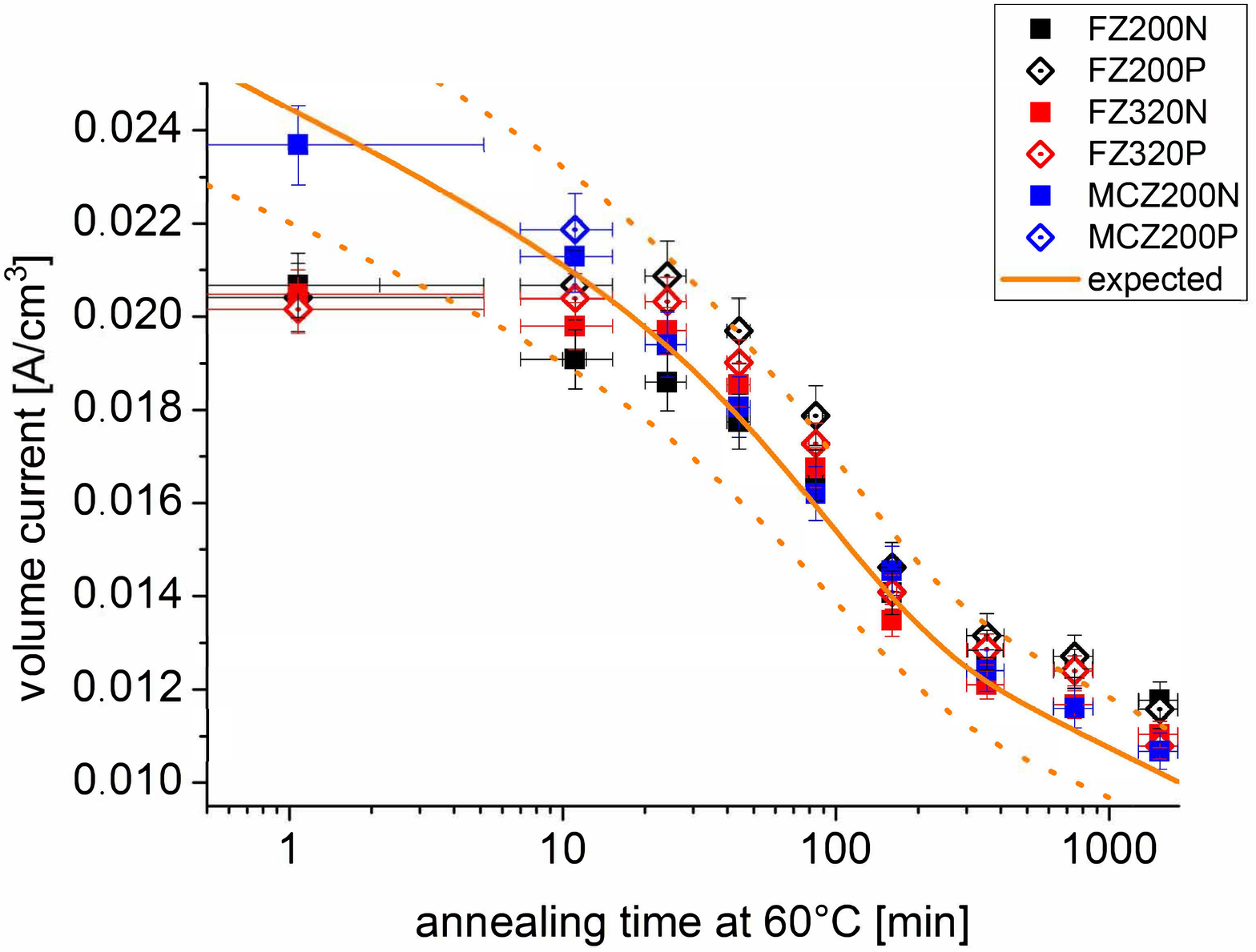}
\caption{Volume current density as a function of annealing time at \SI{+60}{\degree C}  shown for diodes irradiated with neutrons 
to $\phi_{\rm eq}= 4 \times 10^{14}$~cm$^{-2}$. 
For comparison, the current density versus annealing time expected for this fluence based 
on a parameterization and parameters obtained by previous fits to data~\cite{bib:Moll, bib:HamburgModel} 
is depicted by a solid line. 
The corresponding uncertainty, shown as dashed lines, is due to the uncertainty in the fluence, estimated to be about 10$\%$.}
\label{fig:CurrentAn}
\end{center}
\end{figure}

Figure~\ref{fig:Current} shows the volume generation current density as a function of particle fluence for diodes irradiated with \SI{23}{MeV}    or \SI{23}{GeV}    protons, 
neutrons, or a mix of protons and neutrons, respectively. 
To obtain reliable 
results, the currents, taken at 5$\%$ above the full depletion voltage, 
are plotted after annealing for 80 minutes at \SI{+60}{\degree C}. 
The currents are measured at \SI{-20}{\degree C}  and scaled to  \SI{+20}{\degree C} by multiplying with a factor of 59 according to the procedure described in Ref.~\cite{bib:tempscaling}.
The leakage currents per volume are found to be proportional to the fluence, and they are in agreement with 
a proportionality factor $\alpha = 4.1\times 10^{-17}$ A$\cdot$cm$^{-1}$ found in previous measurements, as indicated in the plot.
The measurements confirm that the leakage current universally scales with the NIEL, independent of silicon bulk material and irradiation type.
In turn, the measurements of the leakage current can be used as a cross-check of the particle fluences obtained using dosimetry.
Figure~\ref{fig:CurrentAn} shows the development of the leakage current per volume as a function of annealing time 
at a temperature of \SI{+60}{\degree C} for a variety of diodes irradiated with 
neutrons to $\phi_{\rm eq}= 4 \times 10^{14}$~cm$^{-2}$.  
The leakage current is reduced by about a factor of two after annealing for 1000 minutes at  \SI{+60}{\degree C} and follows the Hamburg model~\cite{bib:HamburgModel}.

\begin{figure}[t]
\begin{center}
\includegraphics[width=0.7\textwidth]{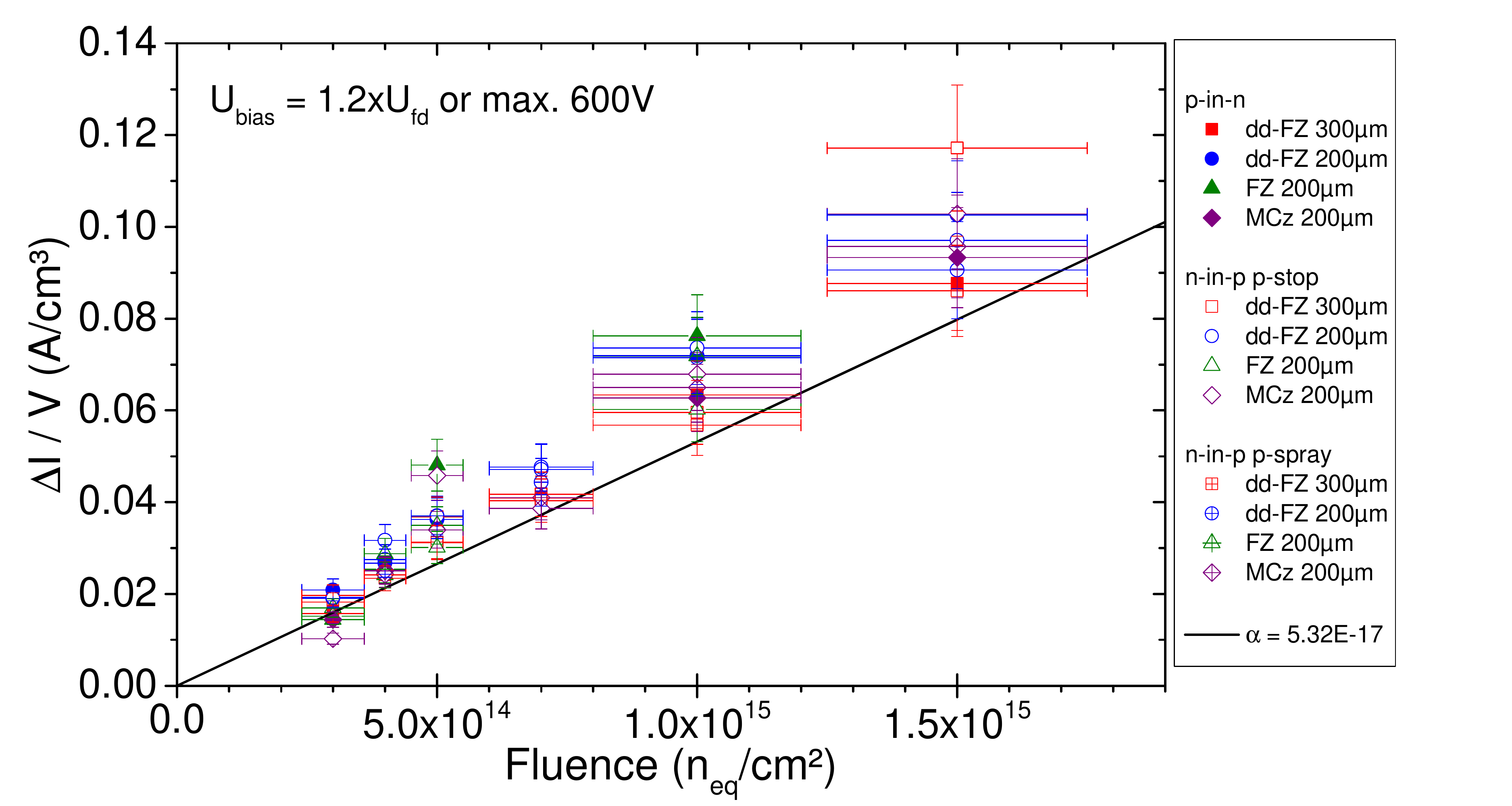}
\caption{Volume current density for strip sensors as a function of particle fluence measured at \SI{-20}{\degree C} and scaled to +\SI{+20}{\degree C}.
The currents were measured after annealing for 10 minutes at +\SI{60}{\degree C} at 20$\%$ above the nominal full depletion voltage, $U_{\rm fd}$.
The data were originally published in Ref.~\cite{bib:ptype}.}
\label{fig:CurrentStrips}
\end{center}
\end{figure}

The current measured for strip sensors 
20$\%$\footnote{For strip sensors, a voltage 20$\%$ above the nominal full depletion voltage was chosen to ensure that the sensors are fully depleted. 
While the change in the volume current above full depletion is very small, an underdepleted sensor would draw less current. 
}
 above the full depletion voltage  after annealing for 10 minutes 
at \SI{+60}{\degree C}  is shown in Fig.~\ref{fig:CurrentStrips}.
For comparison, the straight line describes the volume current as a function of fluence  for diodes with this annealing time ($\alpha = 5.2 \times 10^{-17}$ A$\cdot$ cm$^{-1}$).  
While again a linear dependence on fluence is observed, most sensors display a larger current compared to diodes, 
likely the result of an additional surface current component. 
%
%
\subsection{Full Depletion Voltage}
~\label{sec:depletion}
In this section, systematic studies of the full depletion voltage after irradiation and annealing are presented.
The dependence of the full depletion voltage on particle fluence and type (\SI{23}{MeV}    and \SI{23}{GeV}    protons, neutrons) is studied for 
samples with different silicon crystals, polarity, and thickness.

\begin{figure}[t]
\begin{center}
\includegraphics[width=0.5\textwidth]{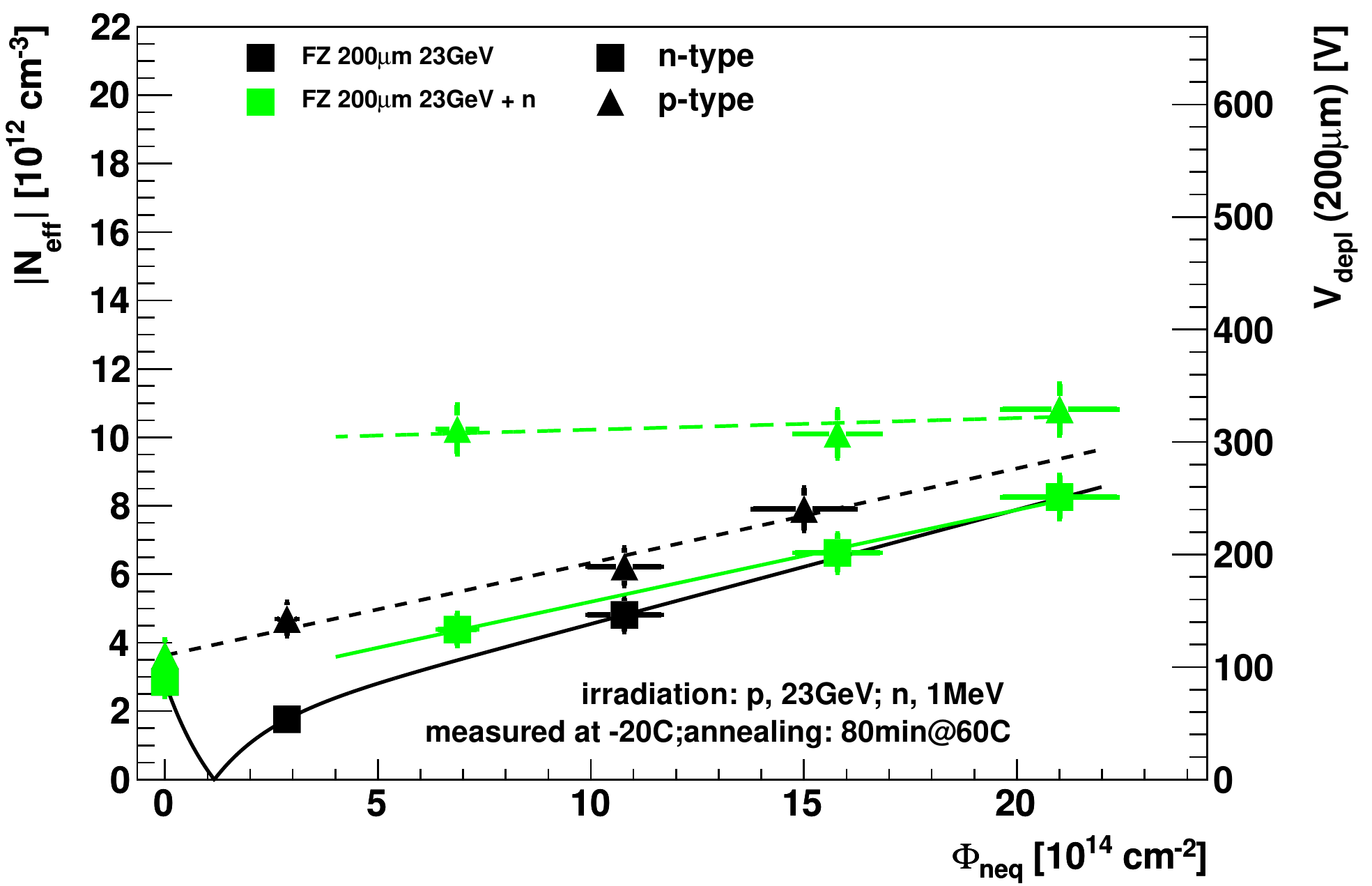}
\caption{
Full depletion voltage and effective space charge concentration determined from capacitance-voltage measurements 
as a function of fluence 
for 200~$\upmu$m thick float-zone diodes for 
\SI{23}{GeV}    proton irradiation (black) and \SI{23}{GeV}    proton plus additional neutron irradiation ("mixed irradiation", green).
The data points are fitted with the Hamburg model~\cite{bib:HamburgModel}. The fit parameters can be found in the appendix of Ref.~\cite{bib:erflethesis}.
}
\label{fig:VdeplFZ}
\end{center}
\end{figure}

Figure~\ref{fig:VdeplFZ} shows the full depletion voltage  and the average effective space charge concentration ($N_{\mathrm eff}$) 
for  \SI{200}{\micro\meter} thick float-zone diodes as a function of 1 MeV neutron equivalent fluence 
for \SI{23}{GeV}    proton irradiation and \SI{23}{GeV}    proton plus additional neutron irradiation ("mixed irradiation").
The effective space charge concentration is related to the full depletion voltage by
\[|N_{\mathrm{eff}}|=\frac{2 \epsilon \epsilon_0}{e}\frac{V_{\rm fd}}{d^2},\]
where $d$ is the thickness of the active volume of the device.

It should be stressed that the concept of an effective space charge concentration constant over the thickness of the sensor derived from the full depletion voltage obtained by $C$-$V$ measurements  has been shown to be inadequate for irradiated sensors in the fluence range studied in this paper ($\phi_{\rm eq} >1\times10^{14}$~cm$^{-2}$ ).  
Irradiated sensors can be better described by a double junction model, which leads to a double peaked electric field~\cite{bib:Eremin, bib:Chiochia}.
In this paper, the full depletion voltage  is merely used as a figure of merit to compare the behavior of different sensors after irradiation and annealing, 
without the same straightforward meaning as  for non-irradiated sensors.

%
The dependence of the full depletion voltage on fluence for the p-in-n float-zone diodes after mixed irradiation is similar to that after proton irradiation only. 
In general, the fact that p bulk material does not undergo space charge sign inversion leads to higher depletion voltages compared to n-type.
For n-in-p devices, the full depletion voltages after mixed irradiation are above the ones for proton irradiation only.
A very large increase in depletion voltage after additional neutron irradiation is visible especially at small fluences.
In this study, the impact of the additional neutron irradiation is larger at smaller fluences as these correspond 
to larger radii in the tracker, which in turn implies a larger neutron fraction 
(Table~\ref{tab:irradiations}). While the cause of this increase might be linked to acceptors created after neutron irradiation, no attempts were made to understand the effect quantitatively.

\begin{figure}[t]
\begin{center}
\includegraphics[width=0.5\textwidth]{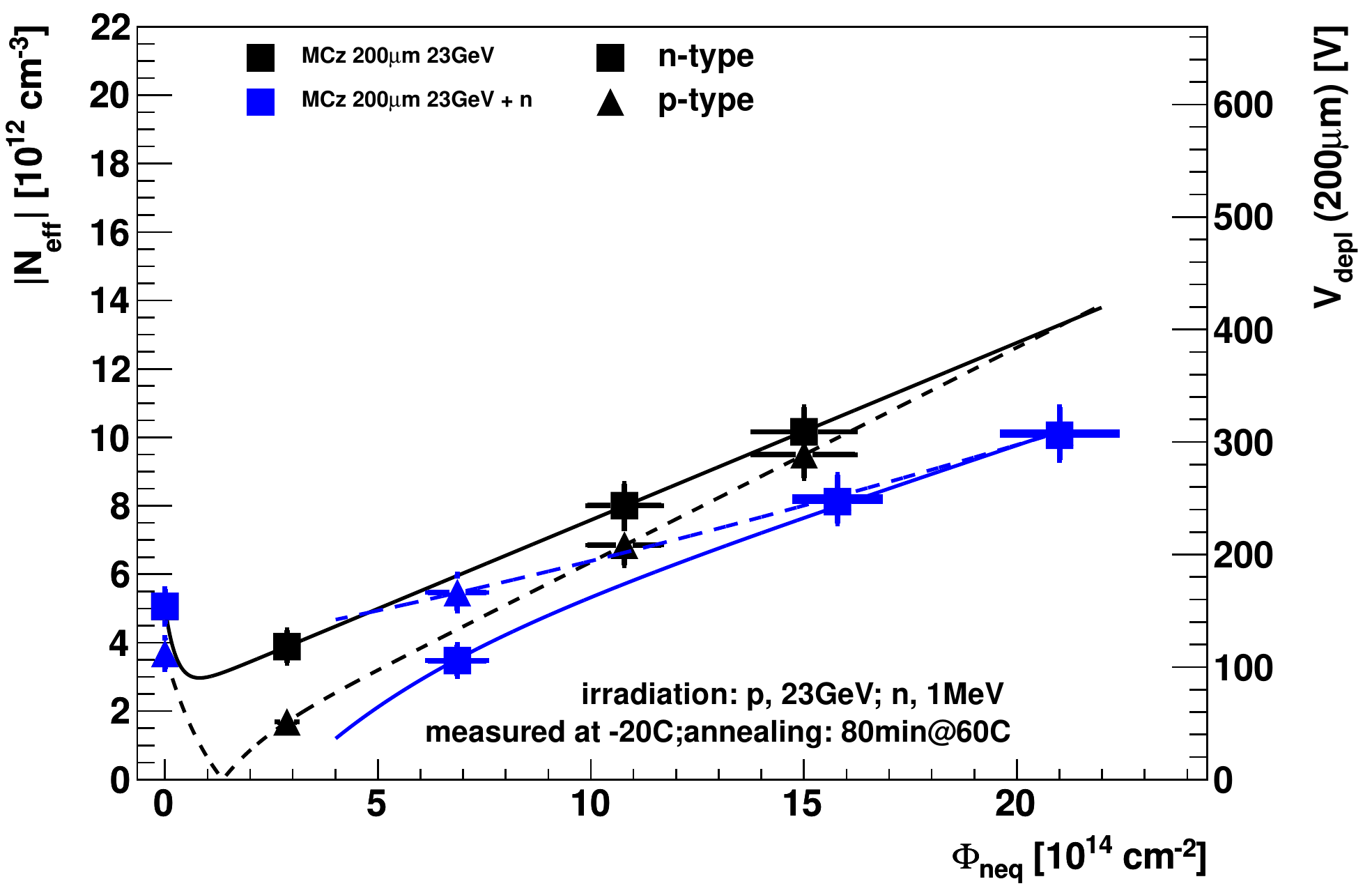}
\caption{
Full depletion voltage and effective space charge concentration determined from capacitance-voltage measurements 
as a function of fluence 
for 200~$\upmu$m thick magnetic Czochralski diodes  for \SI{23}{GeV}    proton irradiation (black) and \SI{23}{GeV}    proton plus additional neutron irradiation ("mixed irradiation'', blue).
The data points are fitted with the Hamburg model~\cite{bib:HamburgModel}. The fit parameters can be found in the appendix of Ref.~\cite{bib:erflethesis}.
}
\label{fig:VdeplMCz}
\end{center}
\end{figure}

Figure~\ref{fig:VdeplMCz} shows the full depletion voltage and effective space charge concentration for  \SI{200}{\micro\meter} thick magnetic Czochralski  diodes as a function of 1 MeV neutron equivalent fluence for \SI{23}{GeV} proton irradiation and mixed irradiation. 
In this case, the full depletion voltages for p-in-n diodes after mixed irradiation lie systematically below the curves after proton irradiation only. 
This can be attributed to donors being created in oxygen rich magnetic Czochralski silicon during GeV proton irradiation, compensating the effect of neutron irradiation-induced acceptors.
These findings confirm that NIEL scaling is violated with respect to the full depletion voltage.
The effects of neutron and GeV proton irradiation partially cancel each other in oxygen rich n-type material, confirming previous 
observations~\cite{bib:Gregor}.

\begin{figure}[t]
\begin{center}
\includegraphics[width=0.5\textwidth]{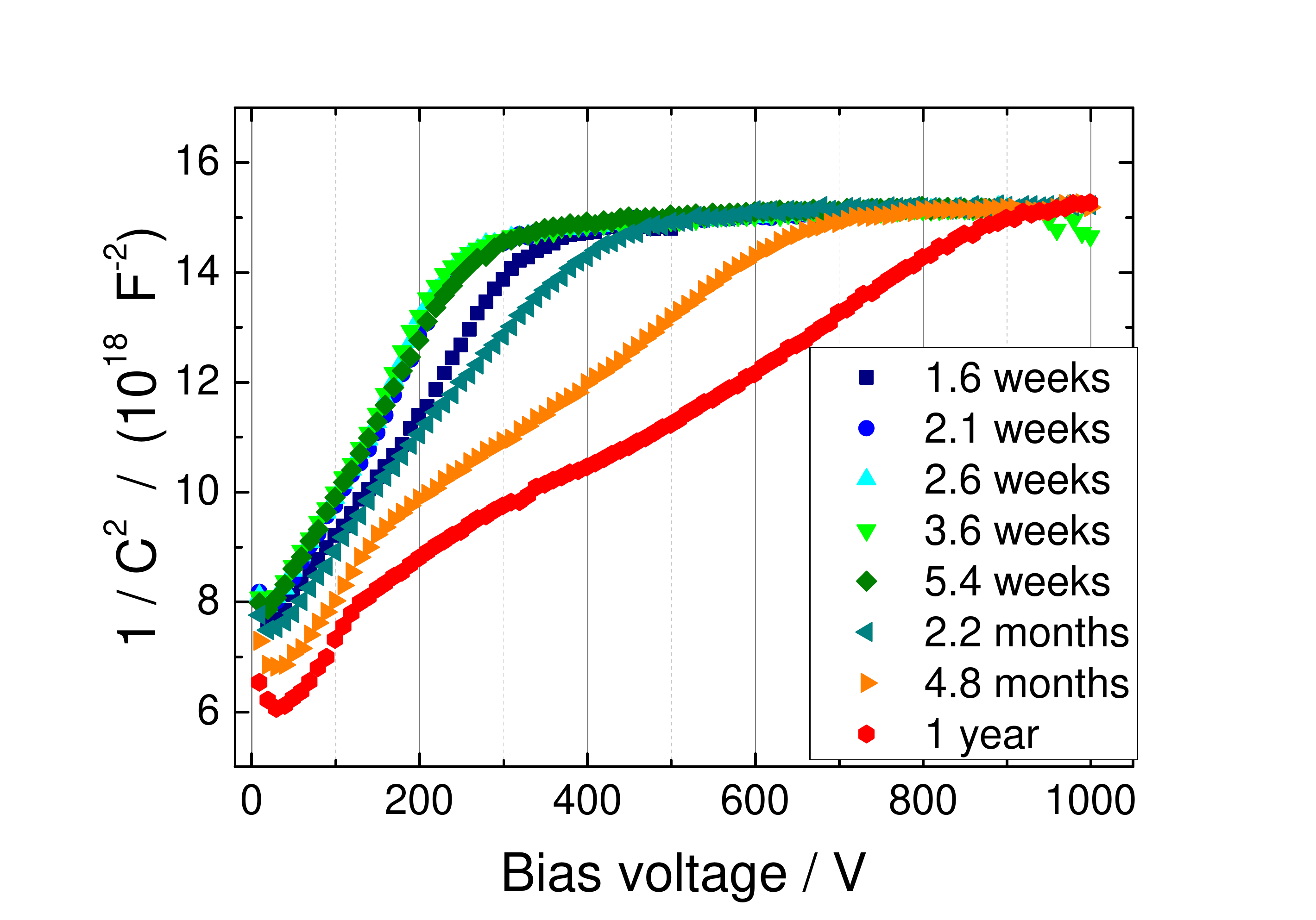}
\hskip -1.1 cm
\includegraphics[width=0.5\textwidth]{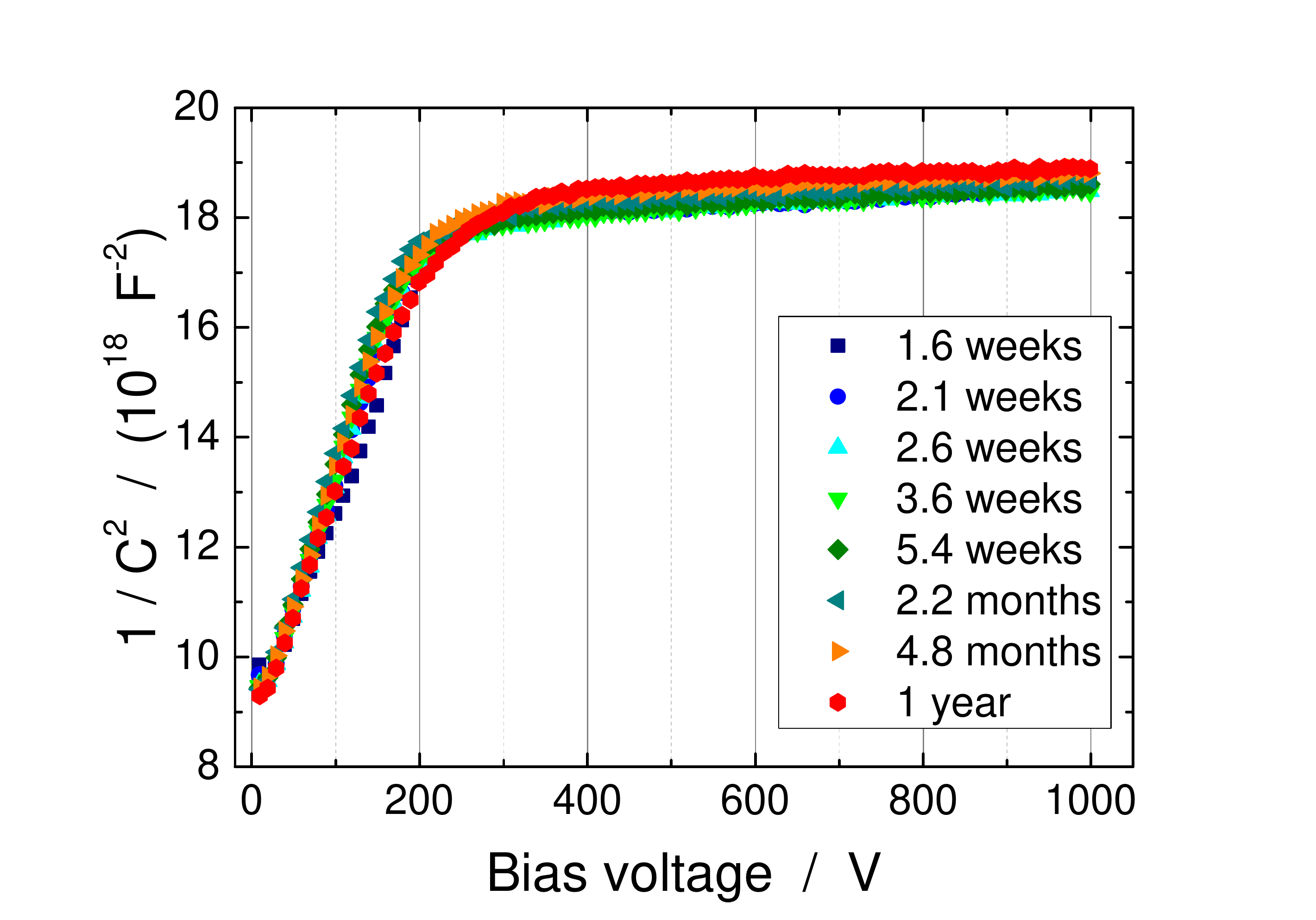}
\caption{Measured inverse capacitance squared ($1/C^2$) as a function of bias voltage for n-in-p float-zone (left) and 
magnetic Czochralski (right) sensors irradiated to $\phi_{\rm eq}=1.5\times10^{15}$~cm$^{-2}$ for different annealing times.
The sensors were annealed at +\SI{+60} or  +\SI{+80}{\degree C}, the times are scaled to the equivalent time at room temperature.
The measurements were performed at \SI{-20}{\degree C} 
and a frequency of 1~kHz.}
\label{fig:CV}
\end{center}
\end{figure}

The differences between float-zone and magnetic Czochralski sensors after mixed irradiation (\SI{23}{GeV}    protons plus 1~MeV neutrons) 
are especially visible in the capacitance-voltage curves.  
In Fig.~\ref{fig:CV}, the measured inverse capacitance squared ($1/C^2$) is plotted  
as a function of applied voltage for 200~$\upmu$m thick n-in-p float-zone 
and magnetic Czochralski mini strip sensors  
irradiated to $\phi_{\rm eq}=1.5\times10^{15}$~cm$^{-2}$ for different annealing times, scaled to room temperature.
While the curves lie virtually on top of each other for the magnetic Czochralski
sensor, a large variation in capacitance below depletion is visible for the float-zone sensor.

\begin{figure}[t]
\begin{center}
\includegraphics[width=0.5\textwidth]{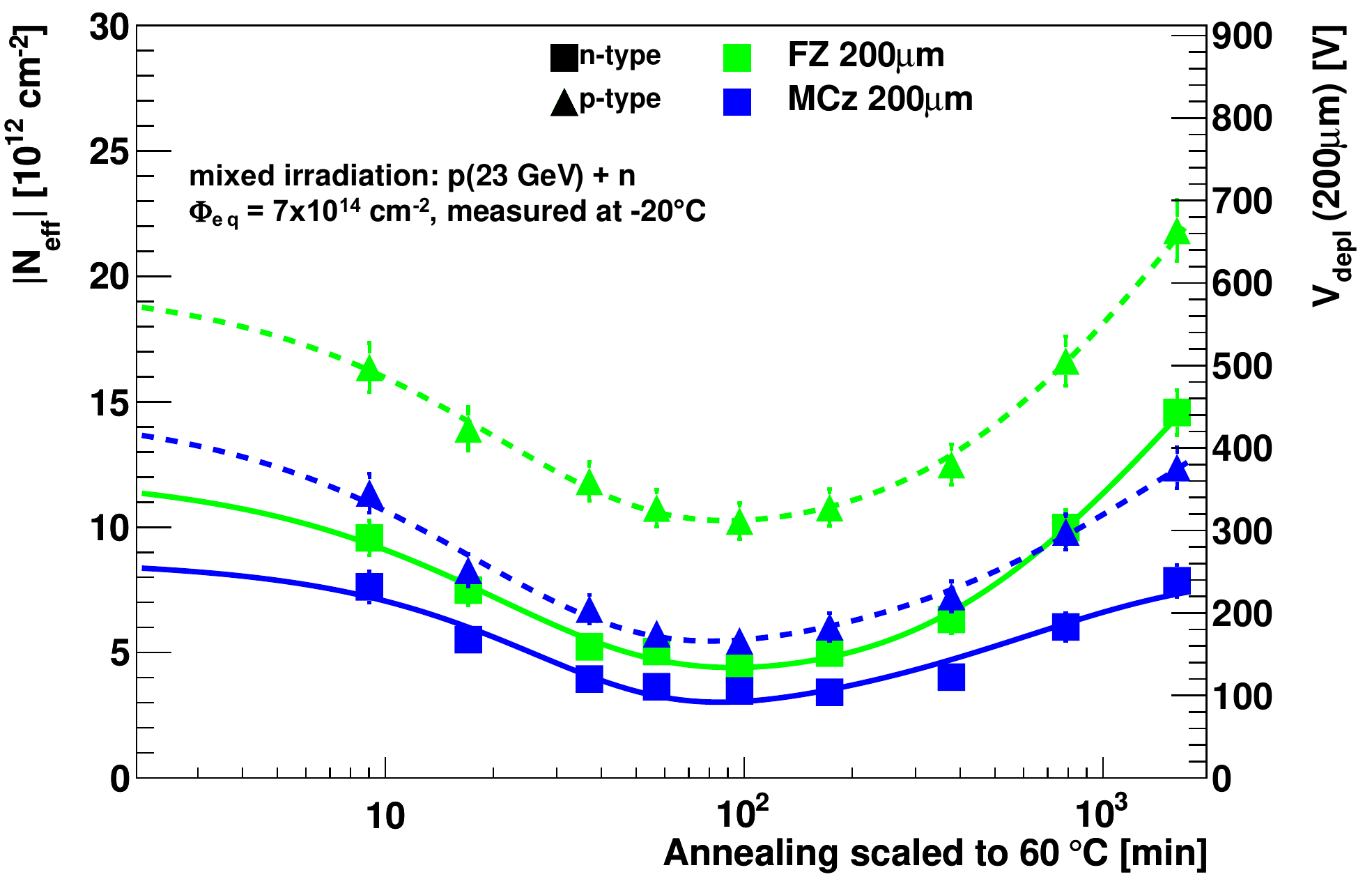}
\includegraphics[width=0.5\textwidth]{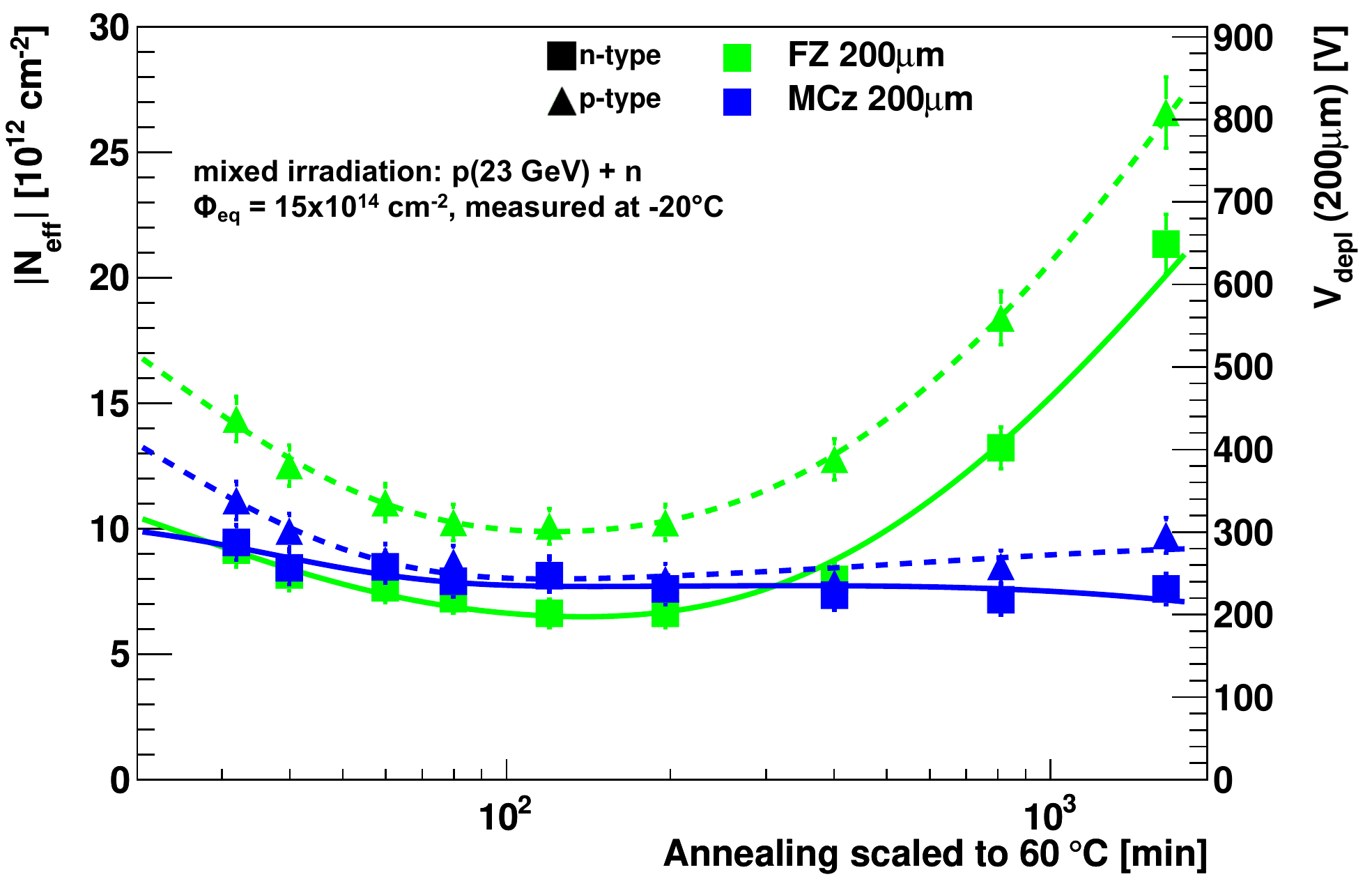}
\includegraphics[width=0.5\textwidth]{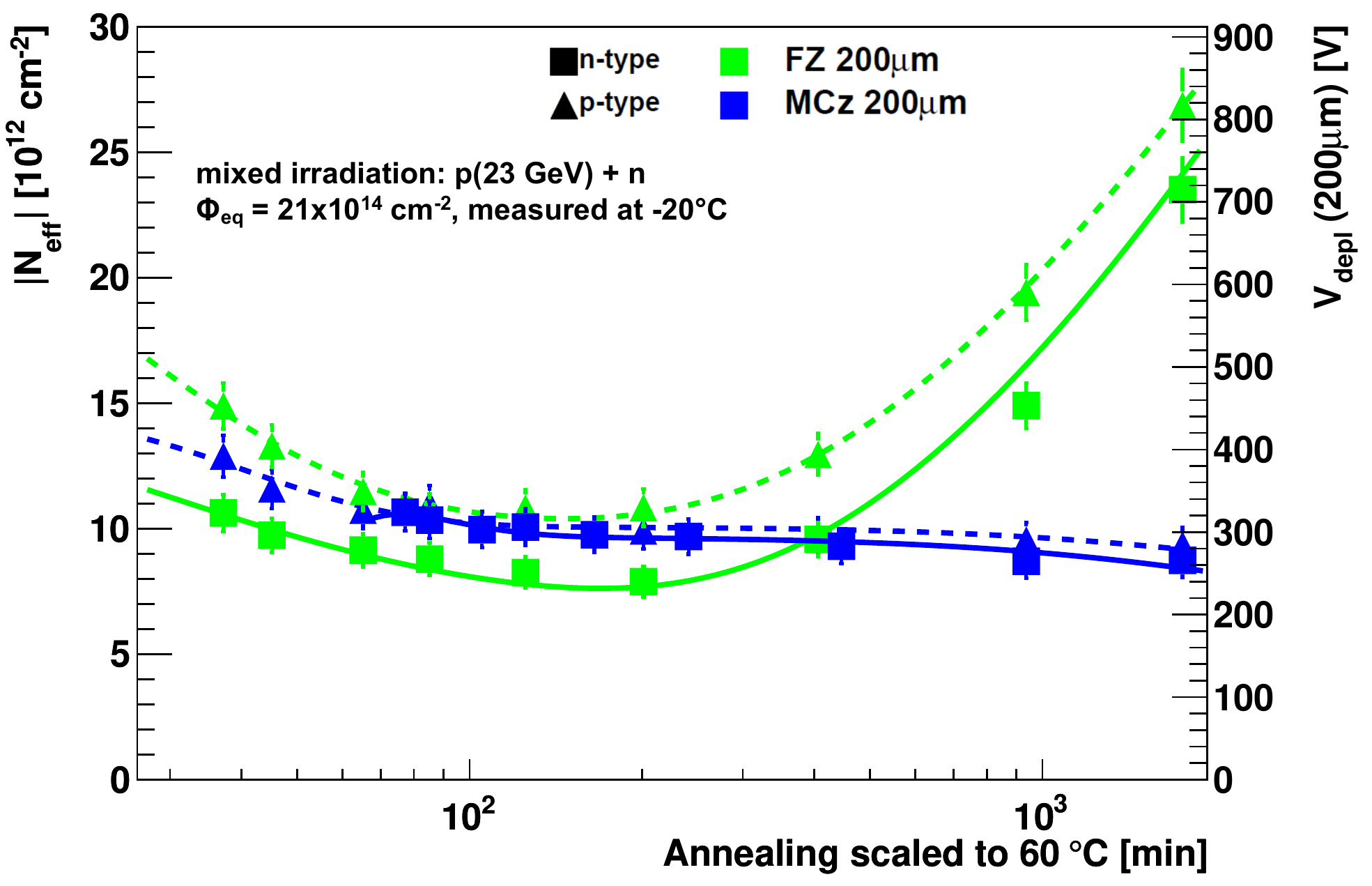}
\caption{Full depletion voltage and effective space charge concentration determined from capacitance-voltage measurements 
after irradiation with \SI{23}{GeV}    protons to $\phi_{\rm eq}=7\times10^{14}$~cm$^{-2}$ (top), $15\times10^{14}$~cm$^{-2}$ (center), and $21\times10^{14}$~cm$^{-2}$ (bottom). 
The full depletion voltage is plotted as a function of annealing time at +\SI{60}{\degree C} for \SI{200}{\micro\meter}  
thick float-zone and magnetic Czochralski diodes.
}
\vskip -0.3cm
\label{fig:NeffAnnealGeV1}
\end{center}
\end{figure}

Figure~\ref{fig:NeffAnnealGeV1} shows the development of the full depletion voltage and effective space charge concentration  
extracted from $C$-$V$ measurements with annealing time after \SI{23}{GeV}    proton plus 1~MeV neutron irradiation 
for $\phi_{\rm eq}=$ 7, 15 and $21\times 10^{14}$~cm$^{-2}$, respectively, for float-zone and magnetic Czochralski diodes.
The annealing time is scaled to +\SI{+60}{\degree C}.
An annealing time of 1000 minutes at +\SI{+60}{\degree C} 
corresponds to 272 days at room temperature (+\SI{+21}{\degree C}), 
based on the temperature dependence of the annealing of the leakage current described in Sec.~\ref{sec:irrad}. 
The data points are fitted with the Hamburg model~\cite{bib:HamburgModel}, the fit parameters can be found in the appendix of Ref.~\cite{bib:erflethesis}.
The full depletion voltages of the n-in-p and p-in-n float-zone sensors show the well-known behavior: an initial drop (short term annealing) to a minimum (stable damage), 
followed by a rise (reverse or long-term annealing).
For the magnetic Czochralski diodes, on the other hand, a very stable full depletion voltage as 
a function of annealing time can be observed, especially for the two larger fluences.
An advantage of the magnetic Czochralski material is that it would be less important to keep the 
tracking detectors cold during maintenance periods to avoid reverse annealing. Moreover, by 
intentionally subjecting the 
sensors to some annealing, the leakage current can be reduced.
%
%
\subsection{Charge Collection}
The hit reconstruction efficiency of the future CMS tracker depends on the collected charge and the 
electronic  noise of each readout channel. For reliable tracker operation, a sufficiently large signal 
that is as stable as possible over the operation time is therefore required.  
For a given sensor, the collected charge depends on 
the fluence it was subjected to, the annealing state, and the applied voltage.  
The goal of this measurement is to study charge collection as a function of these 
parameters and to find a combination of sensor material and thickness of the active 
silicon layer that is suited for sensors for the CMS tracker at the HL-LHC.
To study silicon bulk material effects, 
charge collection is first measured with pad diodes using infrared laser measurements. The study is then extended to 
strip sensors for which the details of the charge collection can be different 
owing to the weighting field and differences in the electric fields.
The weighting field is a measure of the electrostatic coupling between the moving charge and the sensing electrode 
and has units of \SI{}{cm^{-1}}. 

\begin{figure}[t]
\begin{center}
\includegraphics[width=0.5\textwidth]{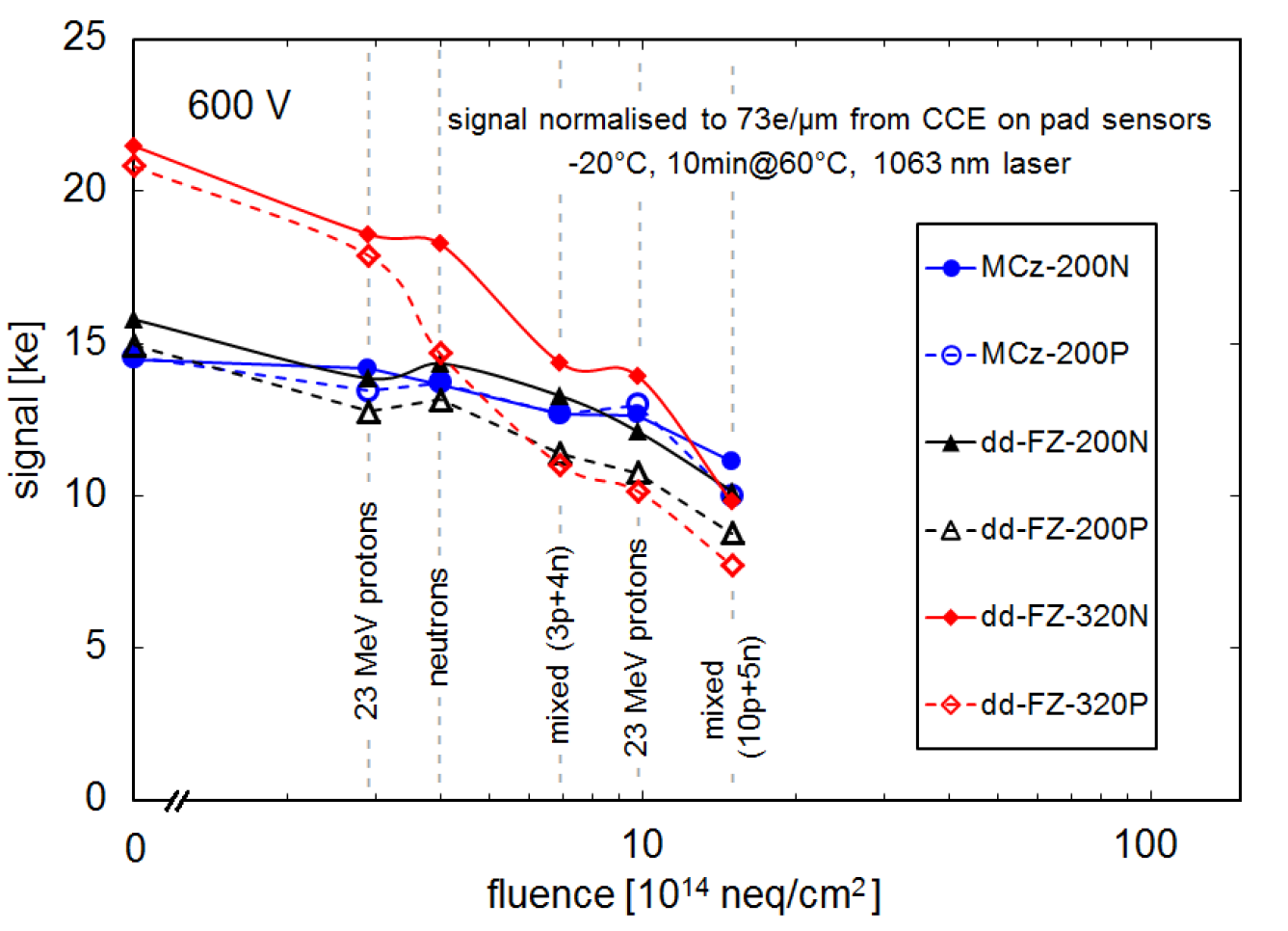}
\hskip -0.3 cm
\includegraphics[width=0.5\textwidth]{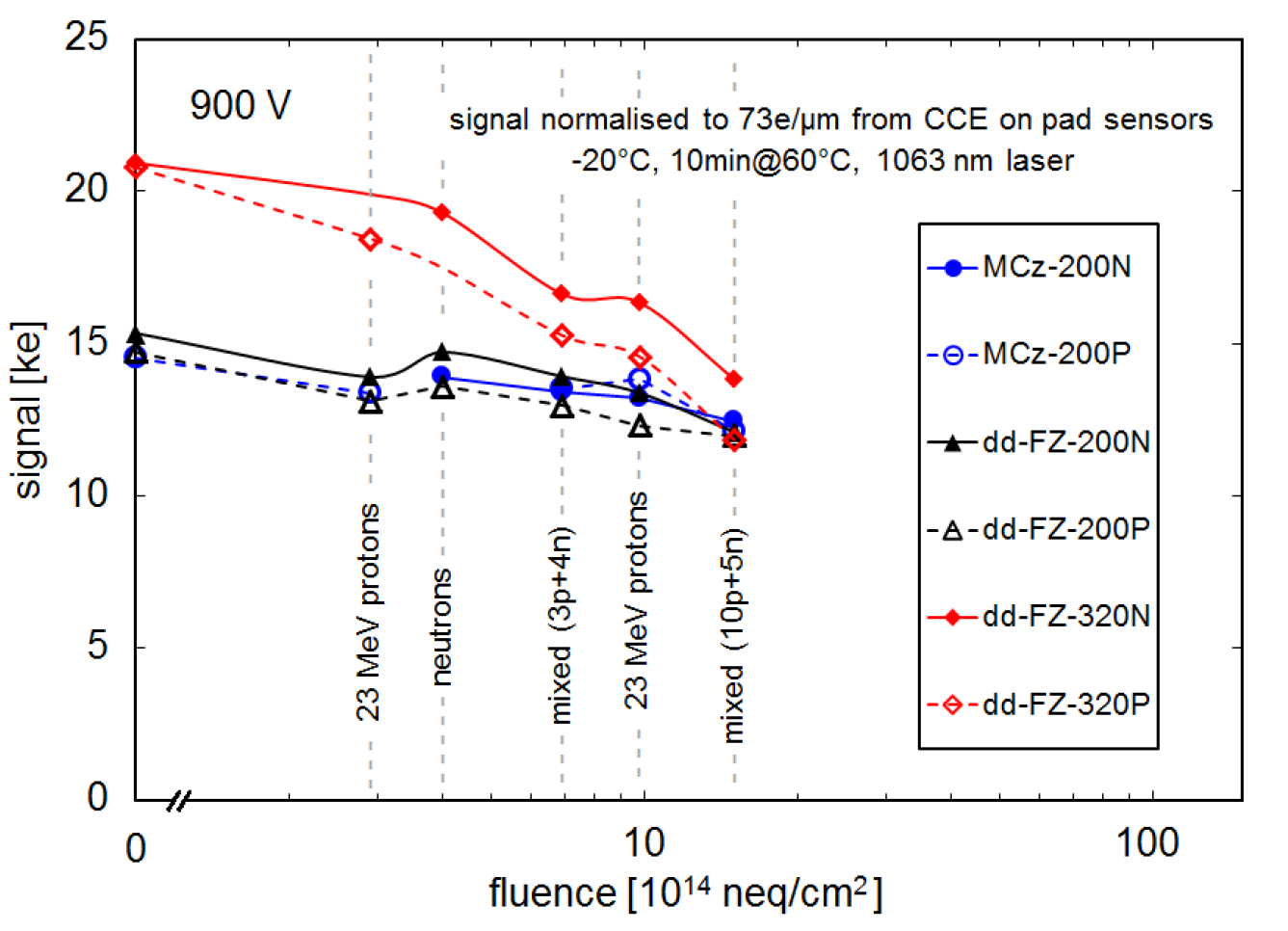}
\caption{Collected charge for pad diodes at \SI{600} (left) and \SI{900}{V} (right), respectively, 
 as a function of fluence for irradiation with 
\SI{23}{MeV}    protons and neutrons, and for mixed irradiadion with $\phi_{\rm eq}=(3+4)\times 10^{14}$~cm$^{-2}$ protons plus neutrons (3p+4n),
and $\phi_{\rm eq}=(10+5)\times 10^{14}$~cm$^{-2}$ protons plus neutrons (10p+5n). The lines are drawn to guide the eye.
The charge collection efficiency, measured by means of an infrared laser, 
has an uncertainty estimated to be around 3\%.
}
\label{fig:chargecol1}
\end{center}
\end{figure}

%
%
First, the collected charge is compared  for 200 and \SI{320}{\micro\meter} thick pad diodes.
Figure~\ref{fig:chargecol1} shows the collected charge 
after irradiation with \SI{23}{MeV} protons and neutrons 
measured at \SI{600} and \SI{900}{V}, respectively. 
The possibility to increase the sensor bias voltage from the nominal \SI{600} to \SI{800}{V} is foreseen 
for the CMS OT at the HL-LHC to obtain larger signals~\cite{bib:Phase2TDR}.
The charge collection efficiency is first  
measured relative to a non-irradiated reference diode using an infrared laser 
and scaled to units of 
collected charge by a factor of  73 electrons per \SI{}{\micro\meter} active bulk silicon; this factor applies to the most probable value of the Landau distribution. 
At lower fluences, more charge is collected in  \SI{320}{\micro\meter} thick silicon than in  \SI{200}{\micro\meter} silicon.
However, for a bias voltage of \SI{600}{V}, the collected charge for the thicker sensors 
drops rapidly as a function of fluence, especially for the  n-in-p deep diffused 
float-zone diode, which 
has been shown to have a full depletion voltage which rises quickly with fluence. 
At \SI{900}{V},  more charge is collected by the \SI{320}{\micro\meter} silicon for all fluences shown.

\begin{figure}[t]
\begin{center}
\includegraphics[width=0.5\textwidth]{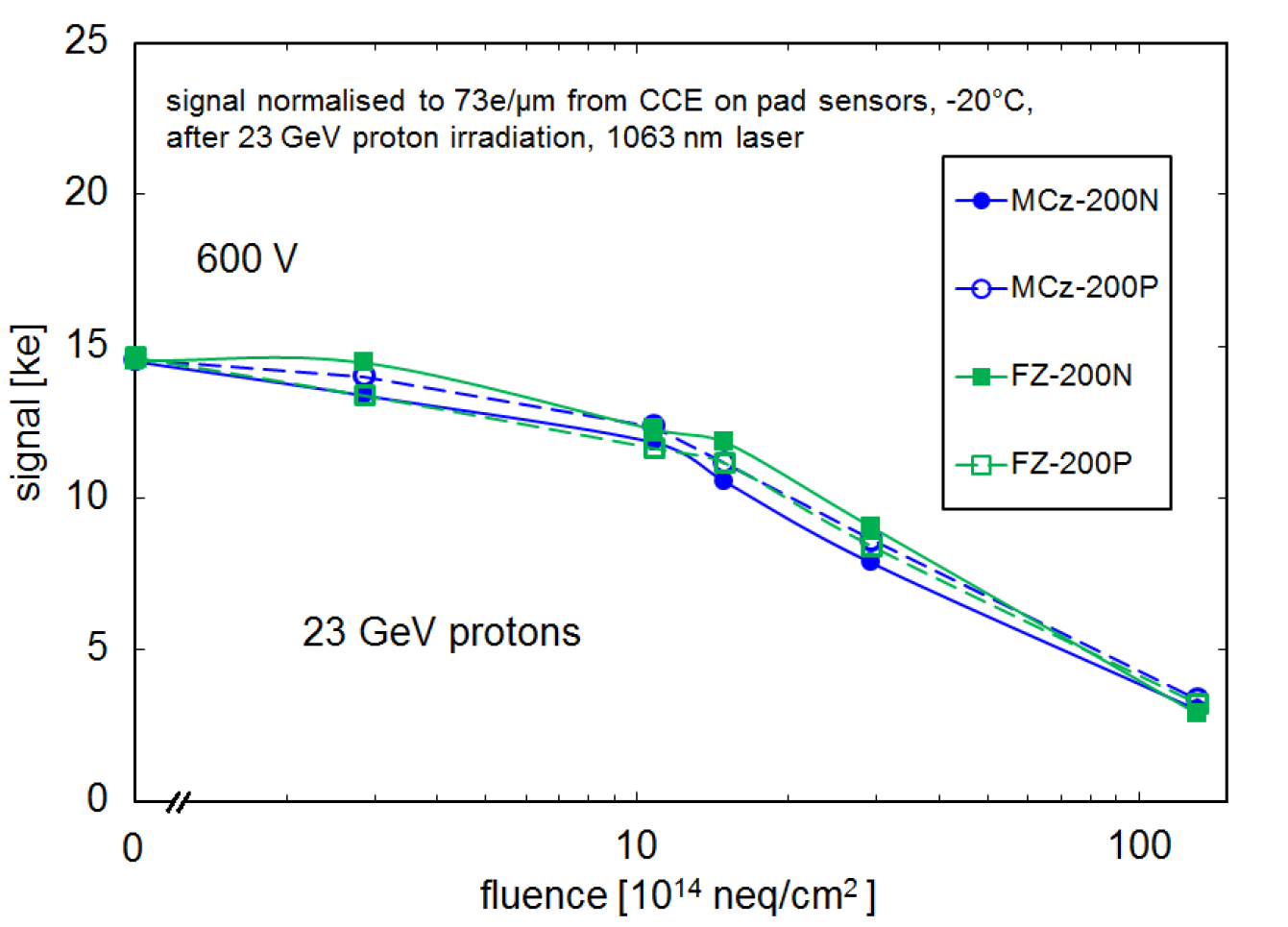}
\hskip -0.3 cm
\includegraphics[width=0.5\textwidth]{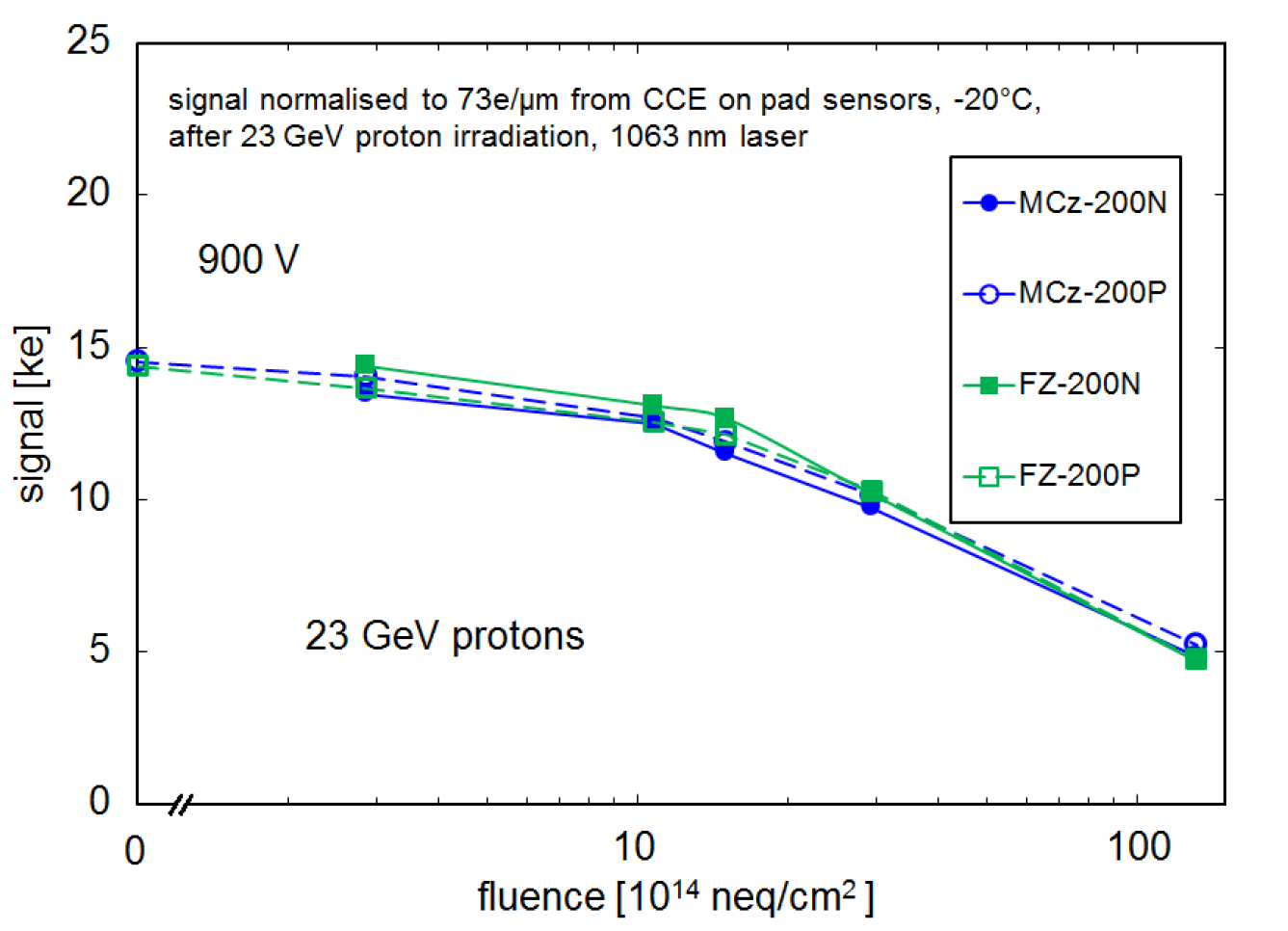}
\caption{Collected charge for pad diodes at \SI{600} (left) and  \SI{900}{V} (right), 
as a function of fluence after irradiation with \SI{23}{GeV}    protons.
The uncertainty in the measured charge collection efficiency is estimated to be around 3\%.
}
\label{fig:chargecol3}
\end{center}
\end{figure}

Next, the collected charge is compared for  \SI{200}{\micro\meter} thick pad diodes 
for different polarities (n-in-p and p-in-n) and bulk materials (FZ and MCz) 
after irradiation with \SI{23}{GeV}    protons. 
The charge collection was again measured at \SI{600} and \SI{900}{V} (Fig.~\ref{fig:chargecol3}).
While at large fluences more signal is collected at \SI{900}{V} bias voltage, very little variation with material and polarity is observed.
All readout chips for the CMS OT at the HL-LHC will feature binary readout~\cite{bib:Phase2TDR}, 
meaning that strips with signals above a threshold are marked and 
only the strip addresses are read out.
The relevant quantity to study is therefore the pulse height of the 
strip with the largest pulse height in a cluster of adjacent strips (``seed charge'') 
rather than the total cluster charge.

The expected noise for the CMS Binary Chip (CBC)~\cite{bib:cbc} used in 2S modules in the outer layers of the OT is of the order of 1000 electrons.
Requiring a threshold of four times the noise, 
and  an MPV of the seed strip three times higher than the threshold leads to a required MPV of 12000 electrons.
For the PS modules in the inner OT layers, the expected noise for the Short Strip ASIC (SSA)  is around 800 electrons~\cite{bib:SSA}, leading to a required threshold of 3200 electrons and a minimum seed signal (MPV) of 9600 electrons.

\begin{figure}[t]
  \centering
 \includegraphics[width=0.67\columnwidth]{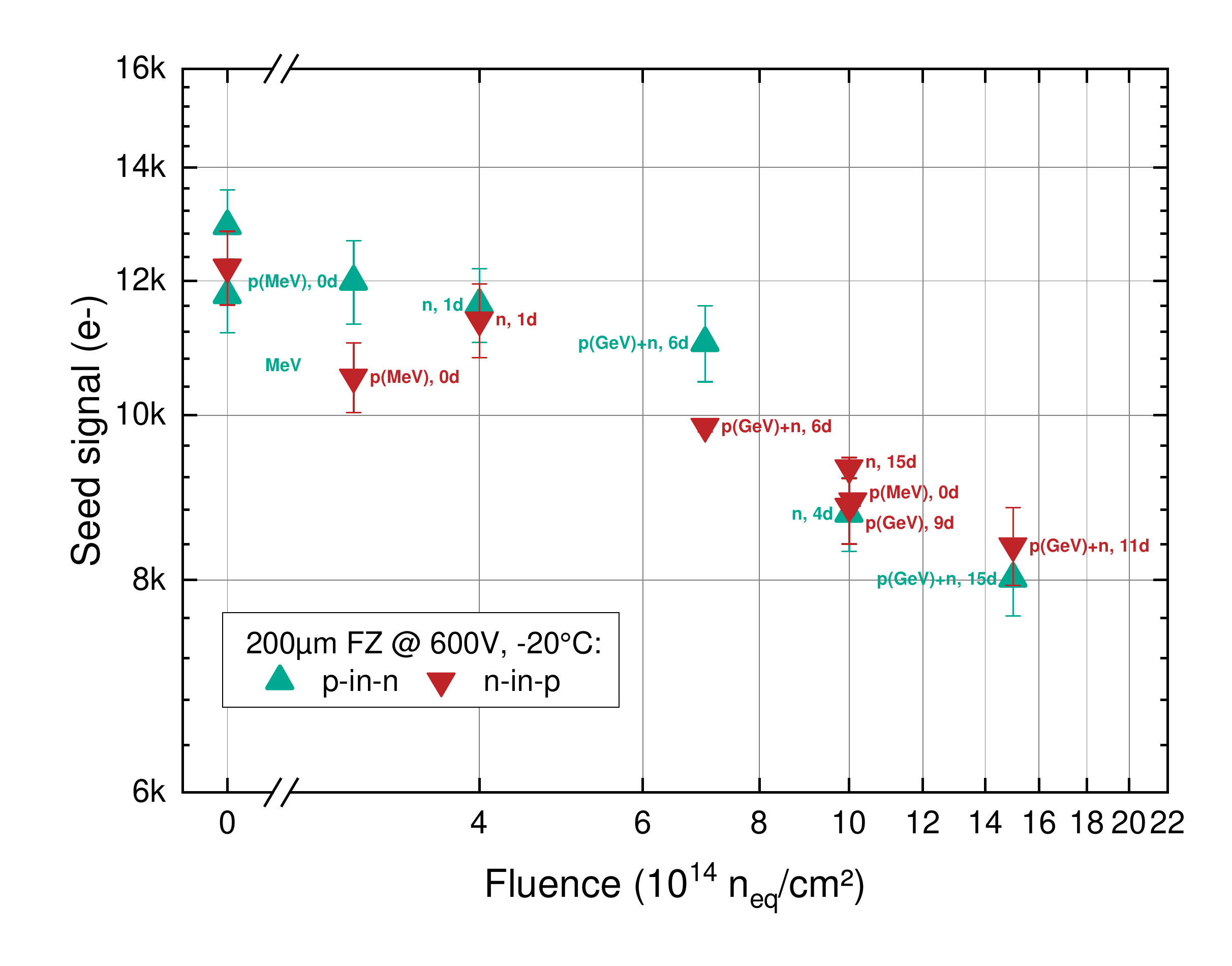}
  \caption[Charge collection \SI{200}{\micro\meter}]{Most probable value of the charge recorded by the seed strip for sensors with a nominal active thickness of \SI{200}{\micro\meter} after annealing  between 0 and 15~days at room temperature. The error bars reflect an estimated error of about 5\% representing statistical and gain uncertainties. The text next to the symbols indicates the irradiation type (\textbf{p} stands for protons with energy range MeV/GeV and \textbf{n} for neutrons). The data were originally published in Ref.~\cite{bib:ptype}.
}
\label{fig:fth200}
\end{figure}

For the following charge collection plots, the ALIBAVA system with a $\beta$ source (Strontium-90) was used.
In Fig.~\ref{fig:fth200} the 
MPV of the charge recorded by the seed strip is shown as a function of particle fluence for 
200~$\upmu$m thick float-zone sensors.
On a log-log scale, the reduction of signal is roughly linear with fluence, 
consistent with a power law. The nominal fluences for an integrated luminosity of \SI{3000}{fb^{-1}} are about $\phi_{\rm eq}=3\times 10^{14}$~cm$^{-2}$ and 
$\phi_{\rm eq}=1\times 10^{15}$~cm$^{-2}$ for the innermost 2S and PS sensors, respectively.
%
For PS as well as 2S sensor modules, the signal for a \SI{200}{\micro\meter}  thick sensor biased to  \SI{600}{V}  would be around or below the required minimum at the highest expected fluence. 
%
%

\begin{figure}[t]
\begin{center}
\includegraphics[width=0.5\textwidth]{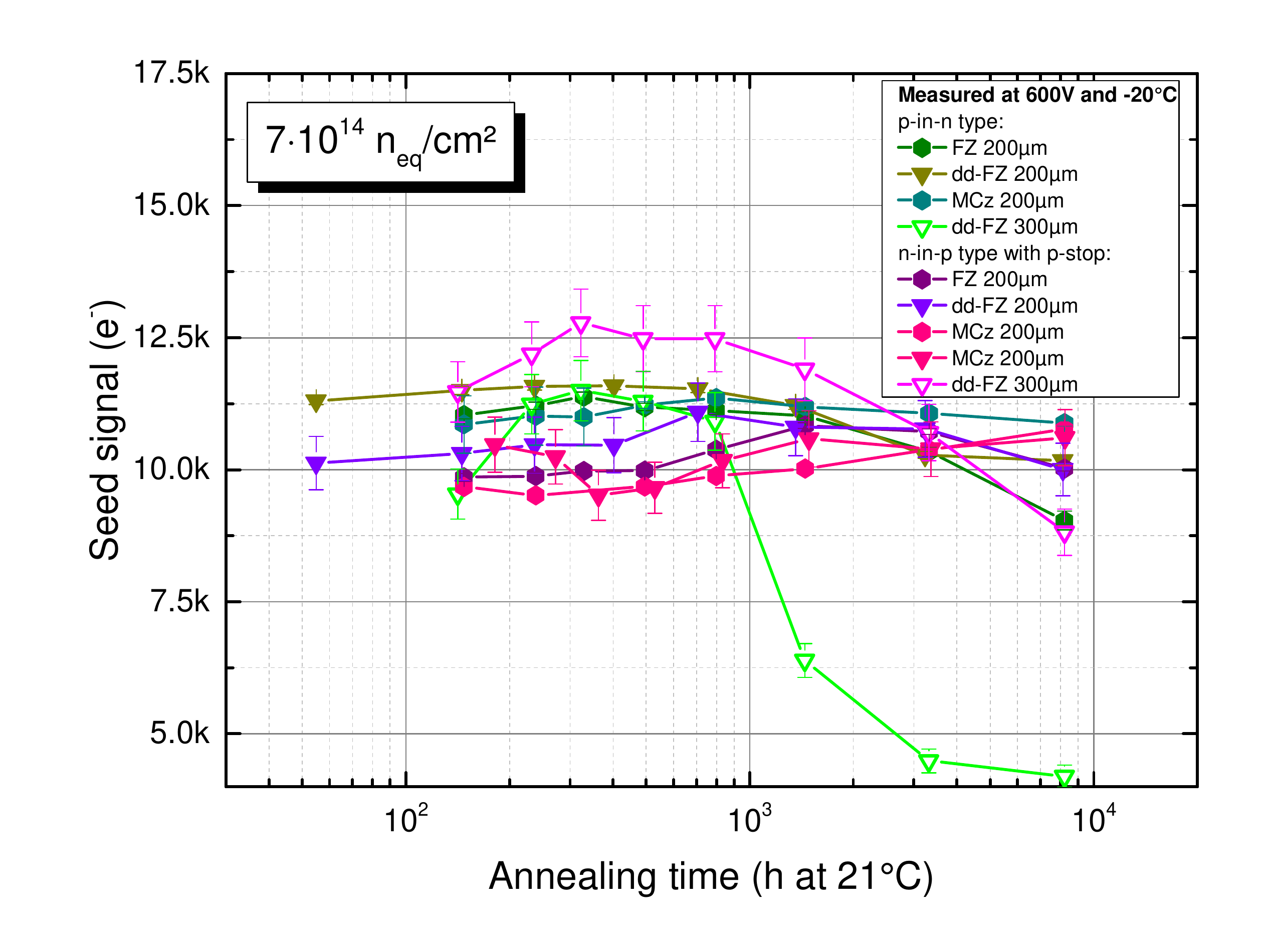}
\hskip -0.3 cm
\includegraphics[width=0.5\textwidth]{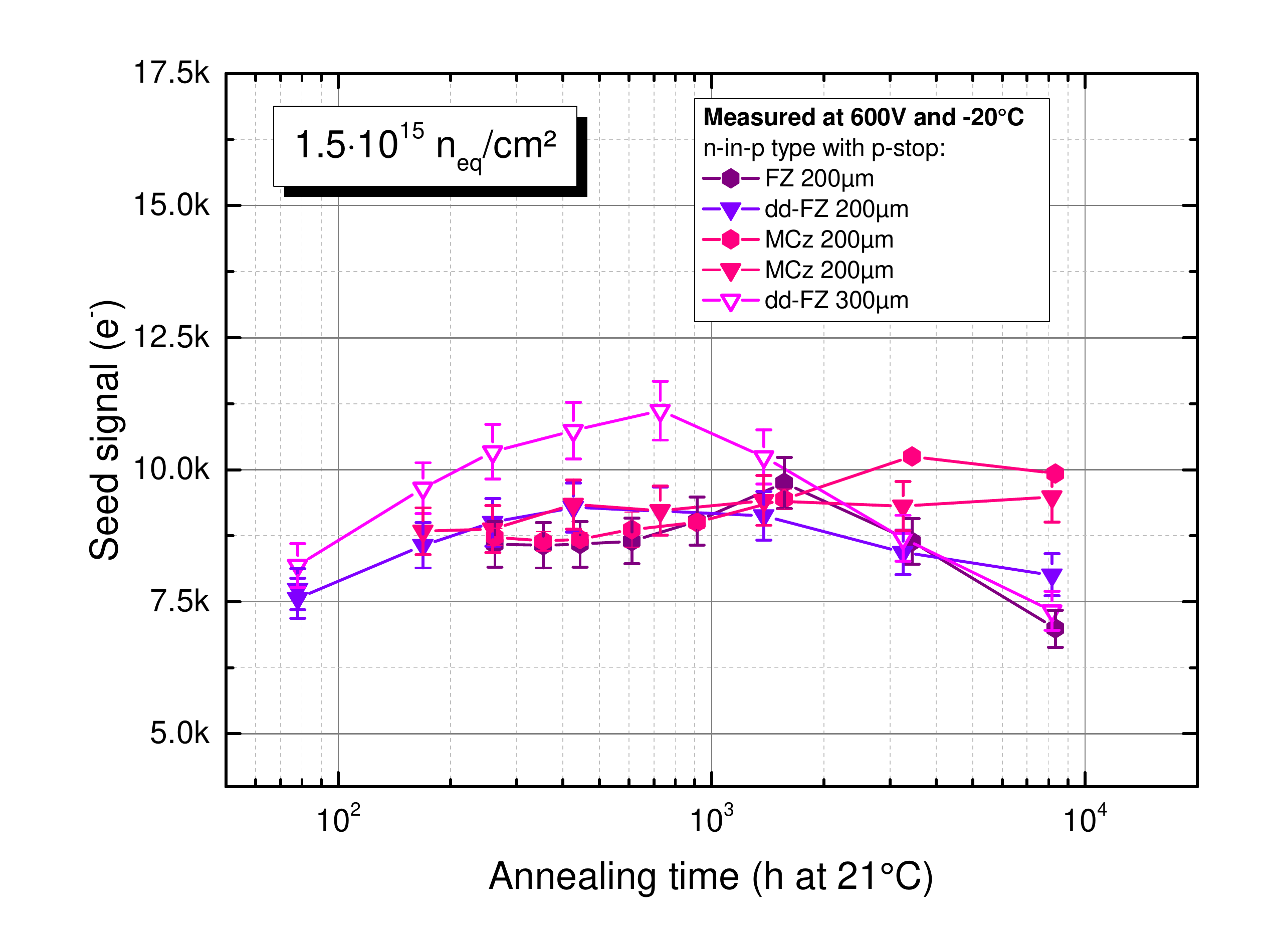}
\caption{
Most probable value of the collected charge for irradiated strip sensors as a function of equivalent 
annealing time for fluences of $\phi_{\rm eq}=$ 7 (left) and 15 $\times 10^{14}$~cm$^{-2}$ (right).  
The sensors were irradiated with a mix of \SI{23}{GeV}    protons and reactor neutrons (hexagonal symbols) and  
\SI{23}{MeV}    protons and reactor neutrons (triangles), respectively. The lines are drawn to guide the eye.}
\label{fig:chargeStripsanneal7E14}
\end{center}
\end{figure}

Figure~\ref{fig:chargeStripsanneal7E14} shows the
MPV of the charge recorded by the seed strip
for  200 and \SI{320}{\micro\meter} thick float-zone and magnetic Czochralski strip sensors as a function of the equivalent annealing time at room temperature after mixed proton and neutron irradiation to fluences of $\phi_{\rm eq}= 7$ and $15\times 10^{14}$~cm$^{-2}$, respectively. 
These fluences correspond to tracker layer radii of \SI{40}{\centi\meter} and  \SI{20}{\centi\meter} 
and an integrated luminosity of \SI{3000}{fb^{-1}}.  The sensors were  biased to \SI{600}{V}. 
The collected charge for the float-zone sensors varies with annealing time, displaying maxima at several hundred hours equivalent annealing time at room temperature and a decrease thereafter.
The decrease is most pronounced at the larger fluence and for the \SI{320}{\micro\meter} thick sensors owing to under-depletion of the sensors.
For the thicker float-zone p-in-n sensor, a very strong decrease of the seed signal with annealing can already be observed at $\phi_{\rm eq} = 7 \times 10^{14}$~cm$^{-2}$,
whereas reliable measurements at $\phi_{\rm eq} = 15 \times 10^{14}$~cm$^{-2}$ were not possible owing to non-Gaussian noise.
The strong dependence of the full depletion voltage on annealing time of the float-zone sensors has already been shown in Fig.~\ref{fig:CV}.
The  magnetic Czochralski sensors, on the other hand, show a very constant charge collection 
as a function of annealing time (see also Sec.~\ref{sec:depletion}).
These findings illustrate again that bias voltages beyond \SI{600}{V} are needed for optimal charge collection for fluences in the range $\phi_{\rm eq} = 7 - 15 \times 10^{14}$~cm$^{-2}$.
%

\begin{figure}[t]
\begin{center}
\includegraphics[width=0.5\textwidth]{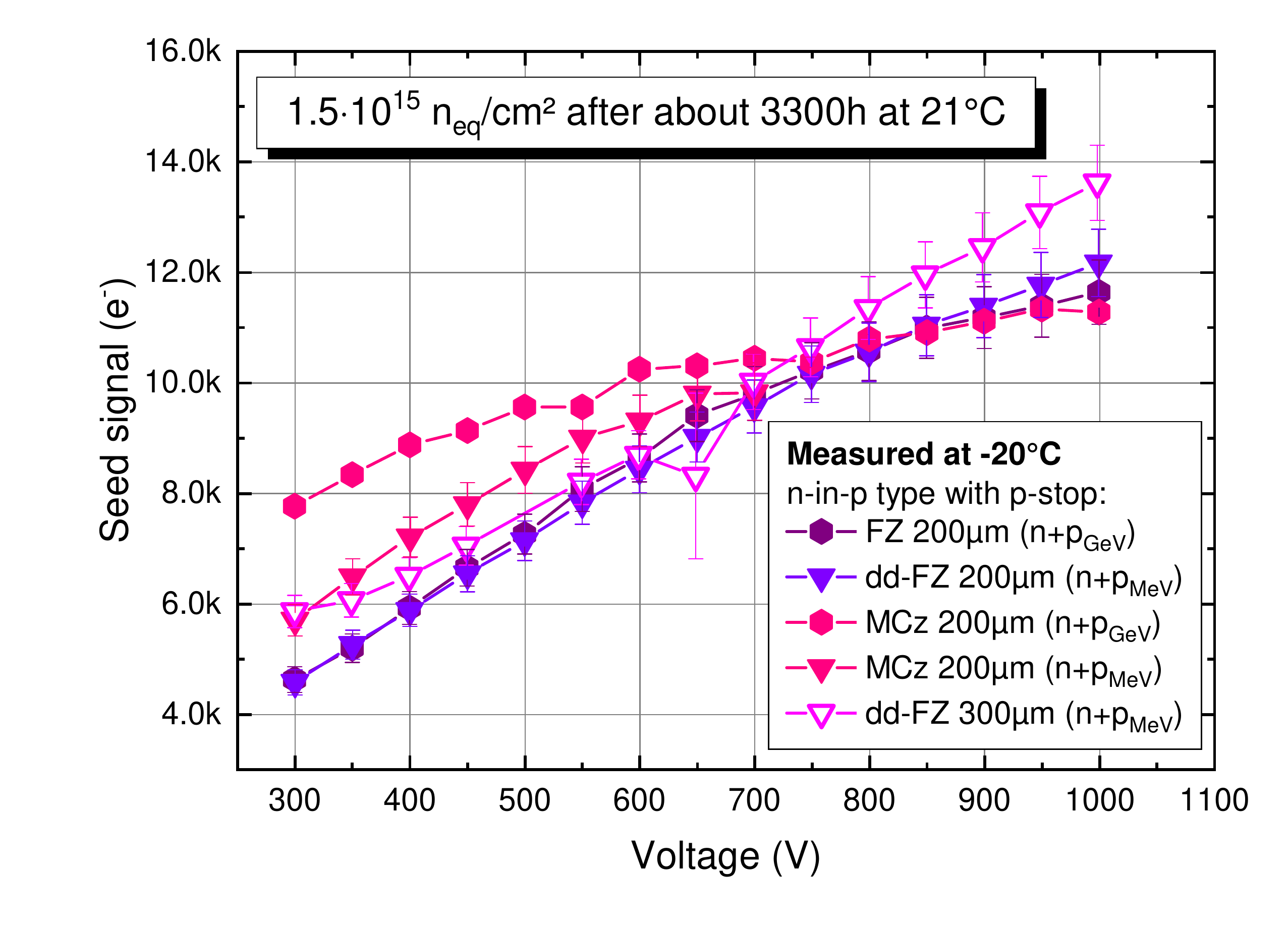}
\caption{Most probable value of the collected charge as function of applied voltage for strip sensors irradiated to $\phi_{\rm eq}=15 \times 10^{14}$~cm$^{-2}$ 
after an annealing time equivalent to 3300 hours at room temperature. The lines are drawn to guide the eye.}
\label{fig:chargeStripsanneal15E14voltage}
\end{center}
\end{figure}

Figure~\ref{fig:chargeStripsanneal15E14voltage} shows the 
MPV of the charge recorded by the seed strip as a function of applied bias voltage for a fluence of 
$\phi_{\rm eq}=15\times10^{14}$~cm$^{-2}$ 
for n-in-p float-zone and magnetic Czochralski strip sensors.
The annealing corresponds to 3300 hours at room temperature. 
%
For voltages up to around 700 V, the seed signal of the \SI{300}{\micro\meter} thick sensor is not higher than that of the 
\SI{200}{\micro\meter} sensors. At higher voltages, thicker silicon leads to additional collected charge under these conditions. 
At voltages up to around 700 V, the  magnetic Czochralski sensor shows an increased charge collection compared to float-zone silicon of the same thickness, 
consistent with the findings described above. At larger voltages, this difference  disappears. 
This again points to the fact that differences in charge collection between materials are due to variations in full depletion voltage and electric field. 
At very large bias voltages, these differences are small and the reduction in charge collection after irradiation and annealing is dominated by trapping, 
which is assumed to be independent of the material. 
The measurements described in this section indicate that $\approx$\SI{300}{\micro\meter} thick sensors would be preferable for the use in the CMS OT with the option to operate them at voltages of up to \SI{800}{V}.

\subsection{Measurements  of Individual Strip Parameters}
\begin{figure}[t]
\begin{center}
\includegraphics[width=0.5\textwidth]{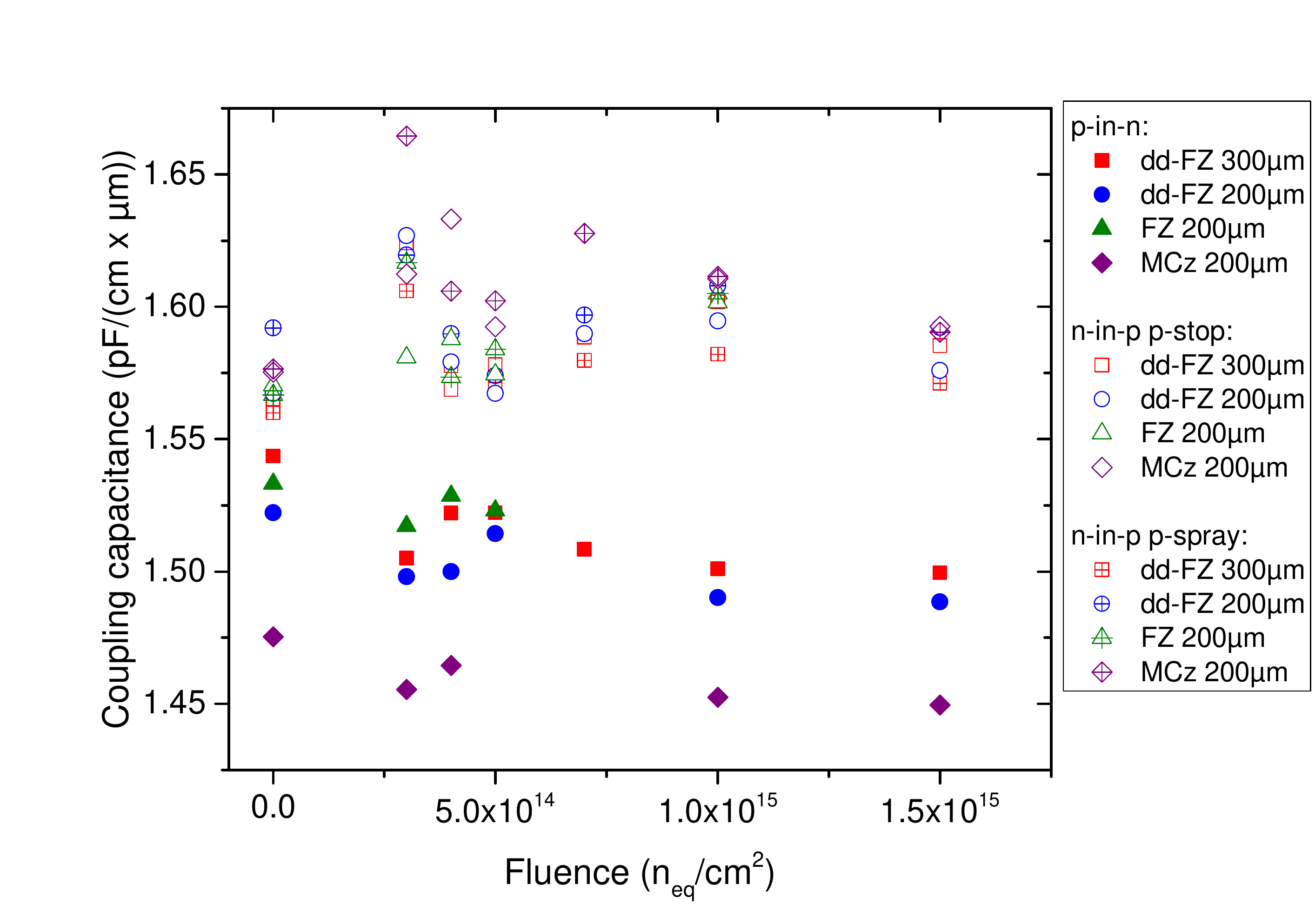}
\caption{Coupling capacitance per strip length and strip width as a function of  fluence. The uncertainty in the measured coupling capacitance is on the order of 2\%.}
\label{fig:strips_CC}
\end{center}
\end{figure}

\begin{figure}[t]
\begin{center}
\includegraphics[width=0.5\textwidth]{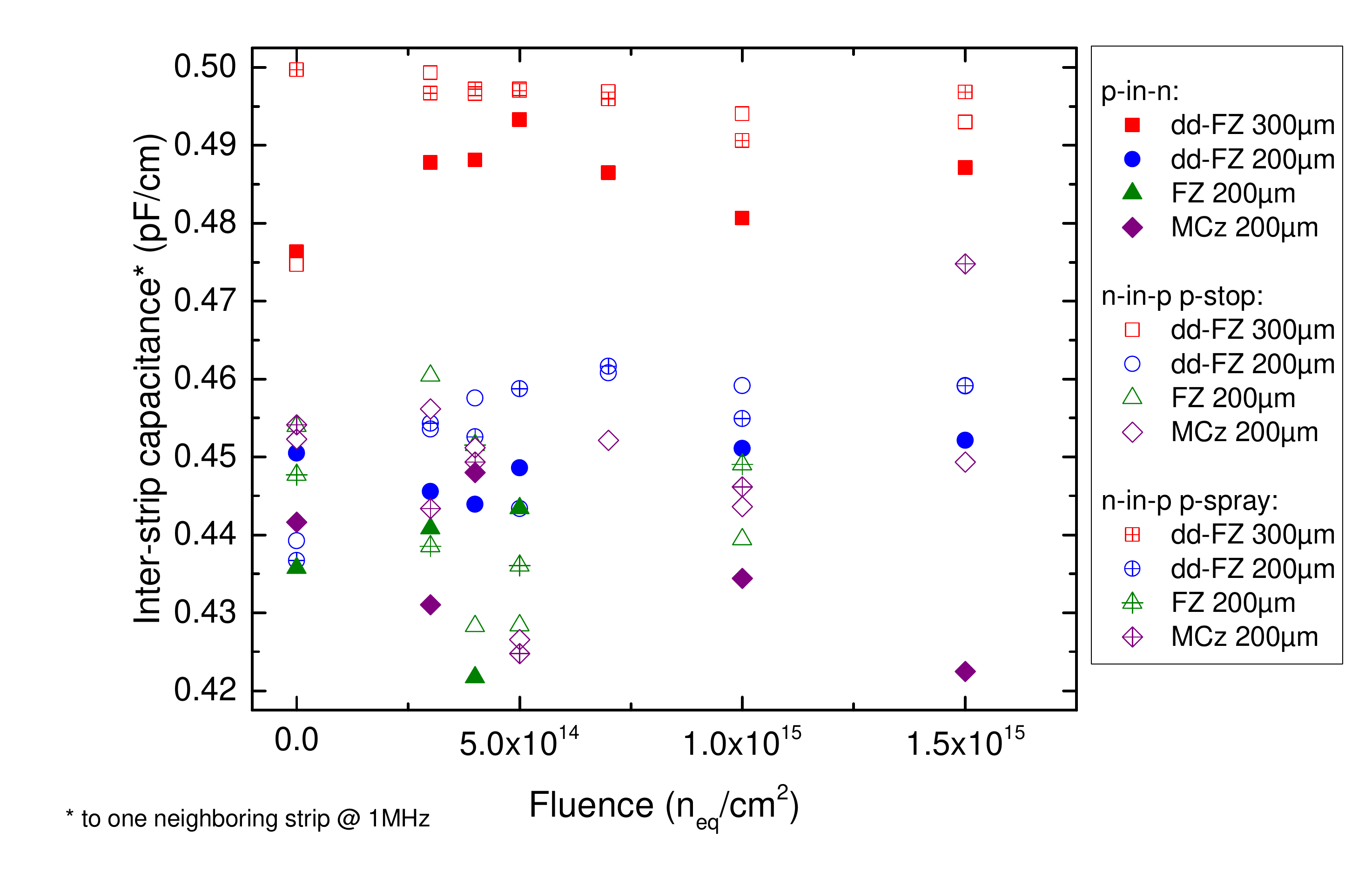}
\caption{Interstrip capacitance to one neighbor per strip length as a function of  fluence. The uncertainty in the measured interstrip capacitance is on the order of 5\%.
The data were originally published in Ref.~\cite{bib:ptype}.}
\label{fig:strips_Cint}
\end{center}
\end{figure}

\begin{figure}[t]
\begin{center}
\includegraphics[width=0.5\textwidth]{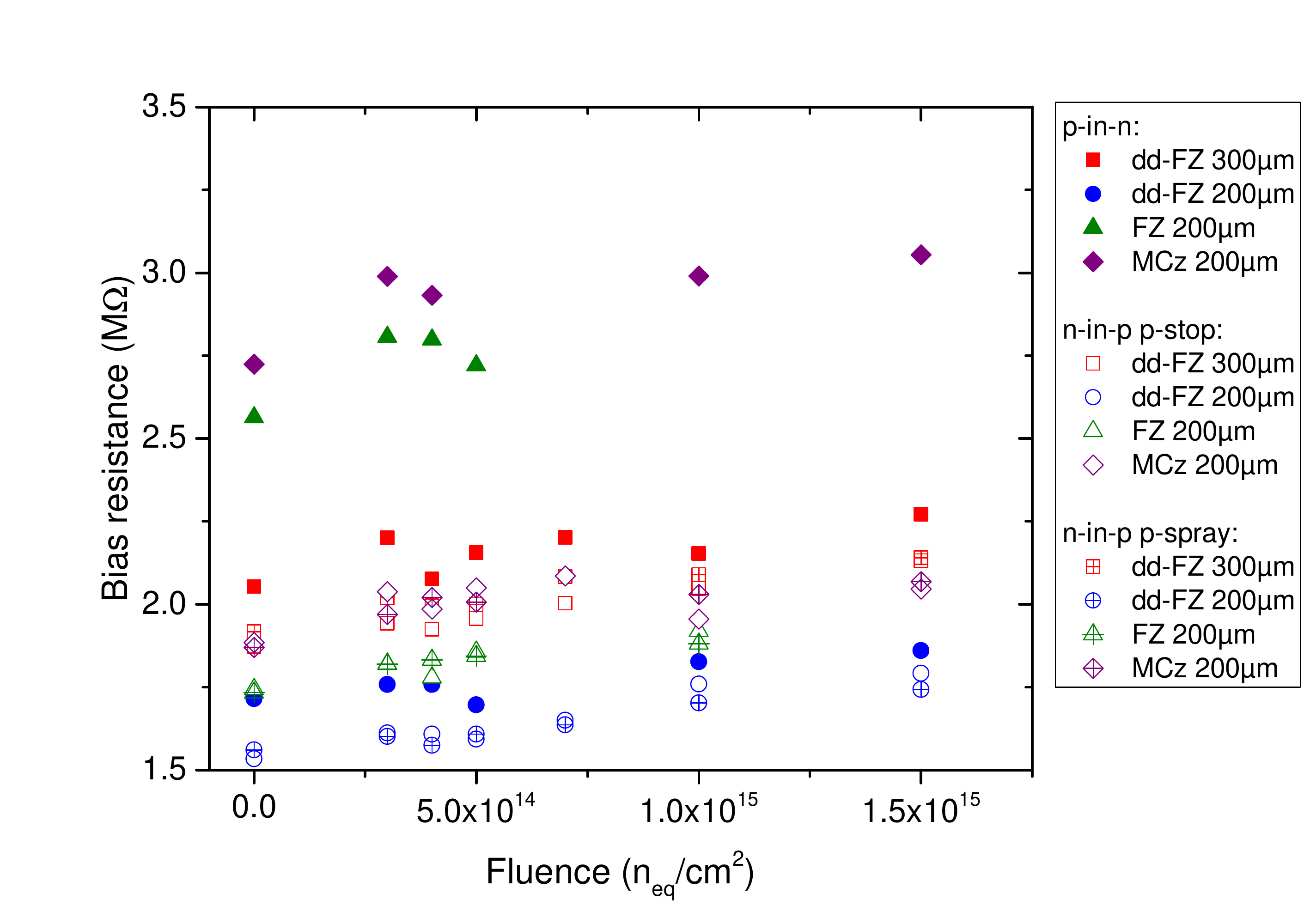}
\caption{Bias resistance as a function of  fluence. The uncertainty in the measured bias resistance is about 1\%.}
\label{fig:strips_Rpoly}
\end{center}
\end{figure}

\begin{figure}[t]
\begin{center}
\includegraphics[width=0.5\textwidth]{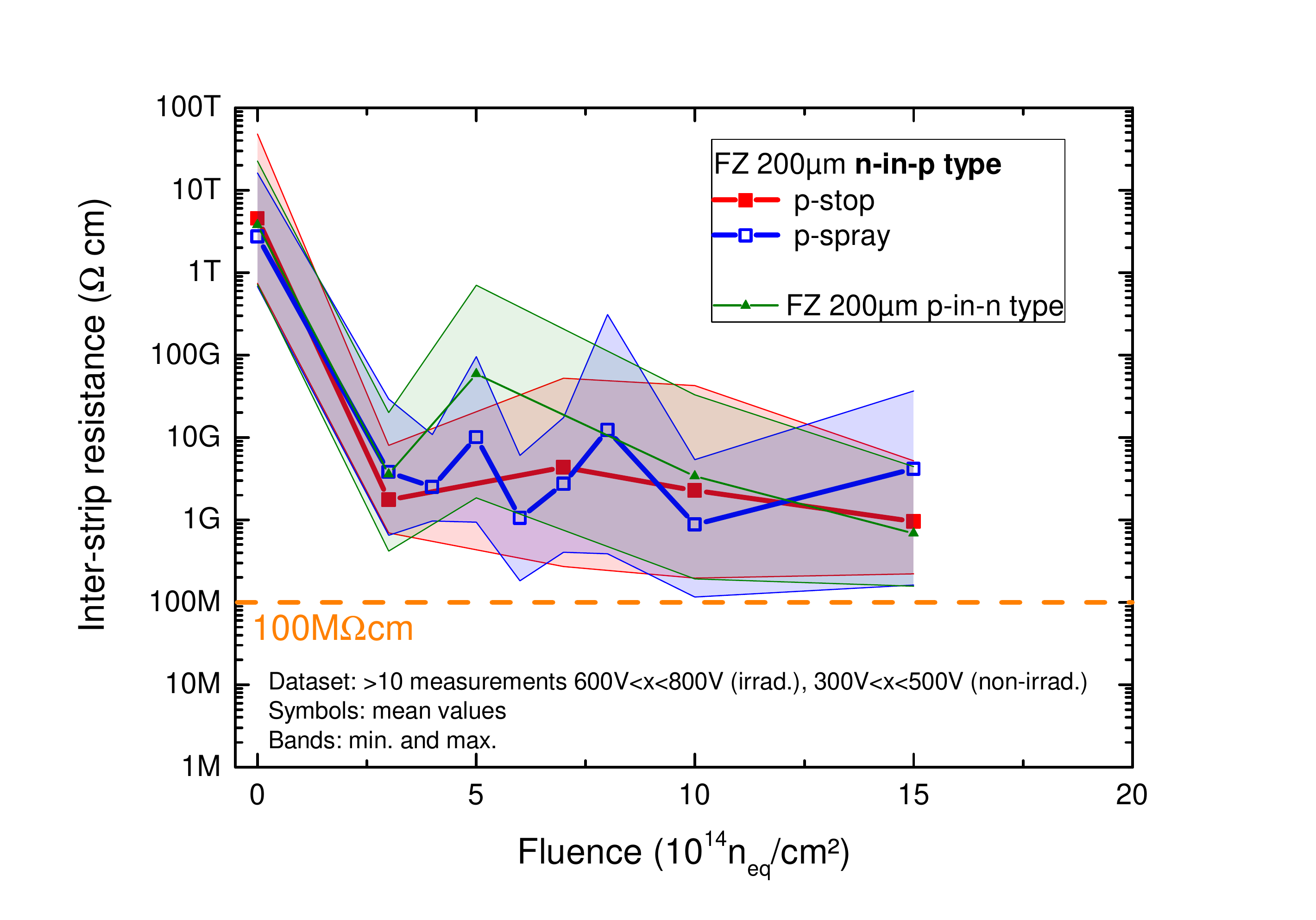}
\caption{Interstrip resistance times strip length as a function of  fluence for \SI{200}{\micro\meter} FZ sensors. The lines are drawn to guide the eye. The uncertainty in the measured interstrip resistance is on the order of 5\%. The data were originally published in Ref.~\cite{bib:ptype}.}
\label{fig:strips_Rint_FZ}
\end{center}
\end{figure}

This section summarizes measurements of strip parameters using probe stations. Details of these measurements and the setups used can be found in Ref.~\cite{Hoffmann_PhD}.\\
The measurements have been performed at +\SI{20}{\degree C} before irradiation and at \SI{-20}{\degree C} after irradiation, after about 10 minutes annealing at +\SI{60}{\degree C} and at a bias 
voltage of \SI{600}{V}. The measurement results as a function of  fluence are shown in Figs.~\ref{fig:strips_CC} --~\ref{fig:strips_Rint_FZ}.
%
%
Except for the interstrip resistance, the measured strip parameters do not show any significant change after irradiations up to 
$\phi_{\rm eq}=1.5\times10^{15}$~cm$^{-2}$. 

The coupling capacitance is shown in Fig.~\ref{fig:strips_CC}. A slightly higher coupling capacitance was observed in n-in-p sensors.
Looking closely, one can observe a small ($\leq 3\%$) increase of the coupling capacitance for n-in-p sensors and a small decrease for p-in-n sensors with increasing fluence. The origin is still unknown, but this effect is very small and does not affect the performance of the sensor at all. The coupling capacitance is specified to be above  
\SI{1.2}{pF/cm}  
per $\upmu$m of implanted strip width.

Figure~\ref{fig:strips_Cint} shows the measurement of the interstrip capacitance. 
The thicker sensors with lower backplane capacitance show higher interstrip capacitance.
The measurement accuracy of the interstrip capacitance is around $5\%$ and within this error no change of this parameter can be observed with increasing fluence. Therefore we do not expect an increasing contribution to the readout noise from the interstrip capacitance, which is specified to be below \SI{0.6}{pF/cm}.

Figure~\ref{fig:strips_Rpoly} shows a step of the measured polysilicon resistance from non-irradiated to the first irradiated samples, which is due to the different temperatures used. The bias resistance increases by about $0.4\%/K$ with decreasing temperature~\cite{Hoffmann_PhD}. The irradiations did not affect the resistance of the polysilicon resistors.
Bias resistances between \SI{1} and \SI{3}{M\Omega}  are acceptable as long as the uniformity across the sensor is good.

The measured interstrip resistance\footnote{The interstrip resistance scales linearly with the inverse of the strip length and the given value has to be divided by the strip length.} 
has been significantly affected by irradiation (Fig.~\ref{fig:strips_Rint_FZ}). It has dropped from a large value above 
1~T$\Omega\cdot {\rm cm}$ to a measured minimal value of 100~M$\Omega\cdot{\rm cm}$ after an irradiation of $\phi_{\rm eq}=1.5\times10^{15}$~cm$^{-2}$. 
No significant difference of p-stop or p-spray isolation is observed for the applied process. 
For a final strip length of 2.5 cm this would result in an interstrip resistance of 40~M$\Omega$, which is still much larger than the bias resistance of about 2~M$\Omega$. 
The measurement is performed by recording the strip leakage current while
applying a small additional potential (\SI{-1}{V} to \SI{1}{V}) resulting in an $I$-$V$ curve
between two strips. The resistance is the inverse of the slope, which
can be quite flat on top of a huge leakage current offset.
The accuracy of this method is not very high and for large currents one can only extract lower limits.
%
\begin{figure}[t]
\begin{center}
\includegraphics[width=0.5\textwidth]{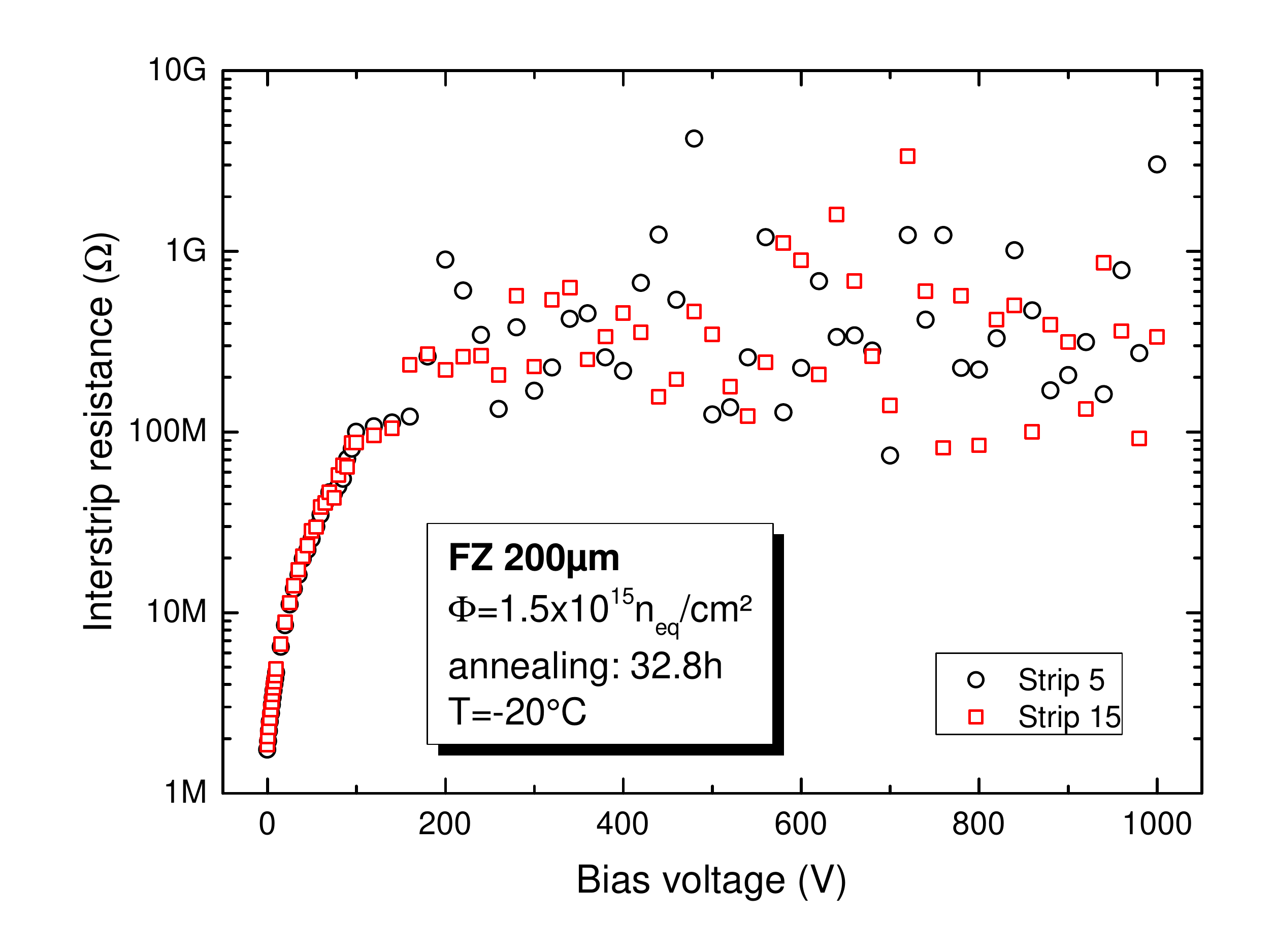}
\caption{Interstrip resistance as a function of  bias voltage for \SI{200}{\micro\meter} FZ sensors.}
\label{fig:strips_Rint_Ramp}
\end{center}
\end{figure}
An example is shown in Fig.~\ref{fig:strips_Rint_Ramp}. At low bias voltages, for which the interstrip resistance is still low, the spread of the measurements is small. For higher resistances the spread increases strongly and the measurements can only reflect the noise; the actual interstrip resistance can be (much) higher. The values in Fig.~\ref{fig:strips_Rint_FZ} are therefore the mean of measurement values between \SI{600} and \SI{800}{V}; the bands reflect the minimum and maximum values. However, the measured interstrip resistance at larger bias voltages is always well above the 
bias resistance, as required for good charge separation for individual strips.

\section{Summary}
CMS has executed a measurement and irradiation campaign to compare silicon sensor materials 
and design choices for Outer Tracker (OT) sensors for the high-luminosity phase of the LHC with the aim to provide input for the decision process.
A number of segmented detectors, special purpose test structures, and pad diodes were implemented by a 
single producer on a variety of wafers. 
In this paper we have presented results of measurements of pad diodes and strip sensors.
Results on the leakage current, the full depletion voltage and 
the charge collection efficiency have been shown before and after irradiation with reactor neutrons, and protons of different energies.
The full depletion voltage for irradiated sensors is simply used as a figure of merit for comparison 
without the same straightforward meaning as  for non-irradiated sensors.

Three silicon sensor materials were studied in detail:  magnetic Czochralski, float-zone, and 
float-zone silicon in which the active thickness was reduced by deep diffusion of dopants from the backside.
The oxygen-rich \SI{200}{\micro\meter} thick magnetic Czochralski sensors have shown a particularly  stable full depletion voltage and 
charge collection with annealing. 
A partial compensation of the effects of irradiation with neutrons and GeV protons 
on the full depletion voltage was observed in n-type magnetic Czochralski material.
As long as the applied voltage is large enough for a given fluence, the differences in charge collection are small, 
and all of these materials are suitable choices. 

The leakage current measured in strip sensors was significantly higher than that 
measured in diodes, likely due to additional surface currents. This has to be taken into account 
in estimates of the sensor module power consumption and cooling needs, and the noise when designing the CMS tracker.
While individual strip parameters change with irradiation, the changes for the fluences studied have been shown to 
be in a range which has little impact on sensor performance. The interstrip isolation has been studied up to 
an   equivalent fluence of $\phi_{\rm eq}=1\times10^{15}$~cm$^{-2}$ and was found to be sufficient
for all sensor types included in this study.

CMS decided in 2013 to use n-in-p sensors for the OT~\cite{bib:ptype, bib:Phase2TDR}.
In addition to the well known advantage of n-in-p sensors of depleting from the segmented side, 
irradiated p-in-n prototype sensors showed strong non-Gaussian noise~\cite{bib:ptype}.

Detailed charge collection studies on diodes and mini strip sensors show that for the entire CMS OT with equivalent fluences up to $\phi_{\rm eq}=1\times10^{15}$~cm$^{-2}$, 
sensors with a thickness in the order of \SI{300}{\micro\meter} are best suited. 
They require operation voltages of up to 800 V at  the end of the HL-LHC running period.
\section*{Acknowledgment}
The tracker groups gratefully acknowledge financial support from the following funding agencies: BMWFW and FWF (Austria); FNRS and FWO (Belgium); CERN; MSE and CSF (Croatia); Academy of Finland, MEC, and HIP (Finland); CEA and CNRS/IN2P3 (France); BMBF, DFG, and HGF (Germany); GSRT (Greece); NKFIA K124850, and Bolyai Fellowship of the Hungarian Academy of Sciences (Hungary); DAE and DST (India); IPM (Iran); INFN (Italy); PAEC (Pakistan); SEIDI, CPAN, PCTI and FEDER (Spain); Swiss Funding Agencies (Switzerland); MST (Taipei); STFC (United Kingdom); DOE and NSF (U.S.A.). \\
Individuals have received support from HFRI (Greece).\\
The research leading to these results has received funding from the 
European Commission under the FP7 Research Infrastructures project AIDA, 
grant agreement no. 262025. 

%








\newpage
\setlength{\parindent}{0pt}
\setlength{\parskip}{6pt plus 2pt minus 1pt}
\newpage
\section*{Tracker Group of the CMS Collaboration}
\label{sec:authorlist}
\addcontentsline{toc}{section}{Tracker group of the CMS collaboration}

\textbf{Institut~f\"{u}r~Hochenergiephysik, Wien, Austria}\\*[0pt]
W.~Adam, T.~Bergauer, D.~Bl\"{o}ch, E.~Brondolin\cmsAuthorMark{1}, M.~Dragicevic, R.~Fr\"{u}hwirth\cmsAuthorMark{2}, V.~Hinger, H.~Steininger, W.~Treberer-Treberspurg

\textbf{Universiteit~Antwerpen, Antwerpen, Belgium}\\*[0pt]
W.~Beaumont, D.~Di~Croce, X.~Janssen, J.~Lauwers, P.~Van~Mechelen, N.~Van~Remortel
% ...

\textbf{Texas~A\&M~University, College~Station, USA}\\*[0pt]
R.~Eusebi

\textbf{Vanderbilt~University, Nashville, USA}\\*[0pt]
P.~D'Angelo, W.~Johns, K.O.~Padeken

\dag: Deceased\\
1: Now at CERN, European~Organization~for~Nuclear~Research, Geneva, Switzerland\\
2: Also at Vienna University of Technology, Vienna, Austria\\
3: Also at Institute for Theoretical and Experimental Physics, Moscow, Russia\\
4: Also at Universit\'{e} de Haute-Alsace, Mulhouse, France\\
5: Also at Horia Hulubei National Institute of Physics and Nuclear Engineering~(IFIN-HH), Bucharest, Romania\\
6: Also at Institut~f\"{u}r~Experimentelle~Kernphysik, Karlsruhe, Germany\\
7: Also at Universit\'{e}~de~Strasbourg, CNRS, IPHC~UMR~7178, Strasbourg, France\\
8: Also at Instituto~de~F\'{i}sica~de~Cantabria~(IFCA), CSIC-Universidad~de~Cantabria, Santander, Spain\\
9: Also at \'{E}cole Polytechnique F\'{e}d\'{e}rale de Lausanne, Lausanne, Switzerland\\
10: Also at Universit\"{a}t~Z\"{u}rich,~Zurich,~Switzerland\\
11: Also at Albert Einstein Center for Fundamental Physics, Bern, Switzerland\\
12: Now at Bethel University, St. Paul, Minnesota, USA \\
13: Now at Karamanoglu Mehmetbey University, Karaman, Turkey\\
14: Now at Florida State University, Tallahassee, USA\\
15: Also at Nanjing Normal University, Nanjing, China

\end{document}